\documentclass[fleqn,usenatbib]{mnras}
\usepackage[utf8]{inputenc}
\usepackage[T1]{fontenc}
\usepackage{multicol}

\usepackage{newtxtext,newtxmath}
\usepackage{ae,aecompl}
\usepackage{times}

\usepackage{geometry}
\usepackage{hyperref}

\usepackage{graphicx}
\usepackage{amsmath}

\usepackage[dvipsnames]{xcolor}
\definecolor{CG}{RGB}{0,0,0} 

\newcommand{\CG}[1]{\textcolor{CG}{#1}}

\definecolor{mnrasblue}{rgb}{0.07, 0.39, 0.8}
\hypersetup{
    colorlinks=true,
    linkcolor=mnrasblue,      
    urlcolor=mnrasblue,
    citecolor=mnrasblue
}

\title[\CG{Observing relativistic features in LSS surveys}]{Observing relativistic features in large-scale structure surveys -- \\
I: Multipoles of the power spectrum}

\author[C. Guandalin, et al.]{
 Caroline Guandalin$^{1}$\thanks{caroline.guandalin@usp.br}, Julian Adamek$^{2,3}$, Philip Bull$^{2,4}$, Chris Clarkson$^{2,4}$,
\newauthor L. Raul Abramo$^{1}$ and Louis Coates$^{2}$
\\
$^{1}$Departamento de F\'isica Matem\'atica, Instituto de F\'isica, Universidade de S\~ao Paulo, Rua do Mat\~ao 1371, CEP \CG{05508-090}, S\~ao Paulo, Brazil\\
$^{2}$School of Physics \& Astronomy, Queen Mary University of London, 327 Mile End Road, London E1 4NS, UK \\
$^{3}$Institute for Computational Science, University of Zurich, Winterthurerstrasse 190, CH-8057 Zurich, Switzerland\\
$^{4}$Department of Physics and Astronomy, University of Western Cape, Cape Town 7535, South Africa
}

\date{Accepted XXX. Received YYY; in original form ZZZ}

\pubyear{2020}

\begin{document}
\label{firstpage}
\pagerange{\pageref{firstpage}--\pageref{lastpage}}
\maketitle

\begin{abstract}
Planned efforts to probe the largest observable distance scales in future cosmological surveys are motivated by a desire to detect relic correlations left over from inflation and the possibility of constraining novel gravitational phenomena beyond general relativity (GR). On such large scales, the usual Newtonian approaches to modelling summary statistics like the power spectrum and bispectrum are insufficient, and we must consider a fully relativistic and gauge-independent treatment of observables such as galaxy number counts in order to avoid subtle biases, e.g. in the determination of the $f_{\rm NL}$ parameter. In this work, we present an initial application of an analysis pipeline capable of accurately modelling and recovering relativistic spectra and correlation functions. As a proof of concept, we focus on the non-zero dipole of the redshift-space power spectrum that arises in the cross-correlation of different mass bins of dark matter halos, using strictly gauge-independent observable quantities evaluated on the past light cone of a fully relativistic $N$-body simulation in a redshift bin $1.7 \le z \le 2.9$. We pay particular attention to the correct estimation of power spectrum multipoles, comparing different methods of accounting for complications such as the survey geometry (window function) and evolution/bias effects on the past light cone, and discuss how our results compare with previous attempts at extracting novel GR signatures from relativistic simulations. 
\end{abstract}




\section{Introduction}\label{sec:intro}
The next generation of galaxy surveys -- such as Euclid, VRO/LSST, and SKA -- will be both wide and deep, covering a broad range of redshifts as well as large areas of the sky, therefore mapping out an unprecedentedly large volume of space and time. On the one hand, this will significantly increase the amount of information available for existing types of cosmological analyses, reducing the sample variance uncertainties on observables such as the BAO scale, redshift-space distortions, and the lensing shear power spectrum. On the other hand, the sheer size of these surveys will also allow qualitatively different cosmological observations to be made. In particular, they will be large enough to access modes on the order of the matter-radiation equality scale $k_{\rm eq}$ \citep[e.g.][]{Philcox:2020xbv}, and possibly even up to the comoving horizon scale $k_\mathcal{H} \sim (a H)$. These represent the very largest observable scales in the Universe, where novel observational features of inflationary and gravitational physics arise that cannot be constrained on the smaller scales probed by existing surveys \citep[e.g.][]{liguori2010, Alonso:2015sfa, Alonso:2015uua, Baker:2015bva, Camera:2014dia, Fonseca:2015laa, Raccanelli:2015vla, Gomes:2019ejy, Bull:2018lat}.

On such large scales, corrections to the standard flat-sky/distant-observer approach to modelling effects such as redshift-space distortions emerge \citep[c.f.][]{kaiser1984}, leading to so-called relativistic corrections or relativistic effects. They have been shown to be an important source of systematic error on large scales, especially for a potential detection of the scale-dependent bias in the galaxy distribution that would be caused by primordial non-Gaussianity \citep{Camera:2014dia, Raccanelli:2015vla, wang2020}. This manifests as
an additional $k^{-2}$ scaling in the bias of dark matter tracers \citep{dalal2008}, which comes from nonlinear corrections to the primordial Bardeen potential due to primordial non-Gaussianities of the local type \citep{komatsu2001}. Relativistic terms with similar $k^{-2}$ scalings also become important on comparable scales \citep[e.g. see][]{Alonso:2015uua, abramo2017}, and so an accurate accounting of them is crucial if we are to recover an unbiased estimate of the non-Gaussianity parameter $f_{\rm NL}$ for example.

Relativistic effects are not only a complicating factor, but contain novel information on the nature of gravity in their own right. Within the context of GR, several unique non-Newtonian features emerge due to such effects. For example, \citet{mcdonald2009} has shown that relativistic effects induce odd multipoles to appear in the cross power spectrum of dark matter tracers, a characteristic with no Newtonian counterpart \citep{Bonvin:2015kuc, Gaztanaga:2015jrs, deweerd2020}. This is, by itself, a new cosmological observable allowing us to probe the equivalence principle at cosmological scales via the Euler equation \citep{bonvin2018}, as well as the gravitational redshift effect \citep*{mcdonald2009, bonvin2014a}. Moreover, by bearing a strong dependence on the Weyl potential, this provides an alternative test for theories of gravity, while the dependence on astrophysical parameters like the magnification and evolution biases opens a new window to a better understanding of the LSS.  Other approaches to constraining deviations from GR via the behaviour of the relativistic effects have also been considered, e.g. \citet*{Lombriser:2013aj, Baker:2015bva}.

In this paper, we develop the basic building blocks of an analysis pipeline that is capable of extracting the relativistic effect signatures from large-scale structure data. As previously mentioned, standard LSS analysis techniques often rely on Newtonian assumptions or the distant-observer approximation, and so it is necessary to adapt them in order to account for the relativistic effects. Relativistic effects also introduce additional dependencies on the astrophysical properties of the source galaxy population(s) that must be accounted for, such as the magnification bias and evolution bias. Using a mock dark matter halo catalogue extracted from the past light cone of a fully relativistic $N$-body simulation generated by the  \textit{gevolution}\footnote{\href{https://github.com/gevolution-code}{https://github.com/gevolution-code}} $N$-body code, we show how these complications can be overcome in the case of relatively idealised catalogue data, with a view to later extending our pipeline to more realistic scenarios.

For the sake of simplicity, we focus only on the detection of odd multipoles caused by relativistic corrections to the redshift-space power spectrum. The relativistic effects that arise in the odd multipoles have the advantage of having a leading-order scaling that goes like $\mathcal{H}/k$, making them easier to detect on scales $k \gtrsim \mathcal{H}$ as compared with the $\mathcal{O}(\mathcal{H}^2 / k^2)$ corrections that affect even multipoles. The dipole is the most straightforward to model and detect, and has the advantage of having previously been detected in the two-point correlation function and power spectrum of halos in the RayGal simulation\footnote{\href{https://cosmo.obspm.fr/raygalgroupsims-relativistic-halo-catalogs/}{https://cosmo.obspm.fr/raygalgroupsims-relativistic-halo-catalogs/}} by \citet{breton2019} and \citet{beutler2020}, respectively, at low redshift. This makes it a suitable target for comparison, although we choose to study higher redshifts of around $z \sim 2- 3$ in order to differentiate our paper from these previous works.

This paper is \CG{organised} as follows. In Section \ref{sec:theory}, we review the theory of relativistic effects in the two-point statistics of biased tracers. In Section \ref{sec:sim}, we describe the \textit{gevolution} light-cone simulation used in this analysis. In Section \ref{sec:ps}, we review the fast Fourier transform (FFT) estimator for the power spectrum multipoles and present our results in Section \ref{sec:results}. Finally, we conclude in Section \ref{sec:conclusions}. For the sake of completeness, we also include Appendix \ref{ap:halos}, which explains the details of the halo catalogues derived from the simulated light cone, and Appendix \ref{ap:window}, where we review the standard method to account for the window function and present some additional results from our measurements.

\section{Relativistic effects in the power spectrum}\label{sec:theory}

Contrary to the simplistic view of $N$-body simulations, which give us the three-dimensional positions of objects at a fixed time slice, the true observed quantity in a galaxy survey is the number of dark matter tracers (e.g. galaxies or halos) $N(z,\boldsymbol{\hat{n}})$ in a pixel given by a solid angle $d\Omega$ around a direction $\boldsymbol{\hat{n}} = (\theta,\varphi)$, defined with respect to the observer's line of sight (LOS), and at a redshift bin $[z, z+\mathrm{d}z]$ \citep{bonvin2011, bonvin2014b}. The number overdensity of some tracer $\alpha$ can thus be defined as 
\begin{equation}
    \delta_{\alpha}^{(s)}(\boldsymbol{s}) \equiv \frac{N_{\alpha}(z,\boldsymbol{\hat{n}}) - \bar{N}_{\alpha} (z)}{\bar{N}_{\alpha} (z)} = \frac{n_{\alpha}(z,\boldsymbol{\hat{n}}) - \bar{n}_{\alpha}(z)}{\bar{n}(z)} + \frac{\delta V(z,\boldsymbol{\hat{n}})}{\bar{V}(z)}, \label{eq:delta_obs}
\end{equation}where the equality is obtained by relating the number counts with the number density as $n(z,\boldsymbol{\hat{n}}) \equiv N(z,\boldsymbol{\hat{n}})/V(z,\boldsymbol{\hat{n}})$. In the above equation, $\bar{N}_{\alpha} (z)$ is the selection function of the tracer $\alpha$, obtained by angular averaging over the tracer number count.

The quantities defined in equation (\ref{eq:delta_obs}) are in redshift space, meaning that they are \CG{characterised} by the observed (comoving) coordinates $\boldsymbol{s} = (s,\theta,\varphi)$, with the radial comoving coordinate $s$ being connected to the observed redshift by some cosmological model\footnote{The radial comoving coordinate in redshift space $s$, obtained from the \textit{observed} redshift, should not be confused with the magnification bias $s_{\alpha}$ of some tracer $\alpha$, which will carry a Greek index throughout this work. We also draw the reader's attention to the radial comoving coordinate denoted by $r$ in real space, obtained from the unperturbed (Hubble flow) redshift of a perfect FLRW universe.}. The standard treatment \citep{kaiser1984}, relating the number of sources in a perfect Friedmann-Lema\^{i}tre-Robertson-Walker (FLRW) universe with the truly observed density field via the conservation of number counts, gives rise to the so-called \textit{redshift-space distortions}. This allows us to relate the theoretical predictions in a homogeneous universe with the observed quantities with the addition of departures from the perfect FLRW metric.

In \citet{kaiser1984}, corrections to the angular pair of coordinates $(\theta,\varphi)$ are not considered, and perturbations to the radial coordinate $s$ come solely from the peculiar velocities of the sources. Even though it describes satisfactorily observations limited to subhorizon scales, where the Newtonian treatment is well suited, this is not a truly observed quantity, as it is gauge-dependent. Furthermore, future galaxy surveys and cosmological observations that rely on the largest (near-horizon) scales demand a proper treatment of the LSS clustering. At smaller scales, the improved sensitivity will also hold the potential for a detection of subleading corrections \citep[e.g. see][]{saga2020}.

Relativistic corrections that appear by considering the covariant definition of redshift have been widely developed in the past decade, and became a paradigm to study large cosmological scales. In addition to solving well-known gauge issues manifested at these scales, it accounts for a number of effects with no Newtonian counterpart. For instance, gravitational redshift and lensing effects are concisely included in equation (\ref{eq:delta_obs}), and we refer the reader to equation (3.23) of \citet{yoo2014} and equation (16) of \citet{bonvin2014b} for its full expression.

By collecting the terms proportional to $\boldsymbol{v}\cdot\boldsymbol{n}$ we end up with \citep{bonvin2014b,clarkson2019}
\begin{equation}
    \delta_{\alpha}^{(s)}(\boldsymbol{r}) = b_{\alpha} \delta^{(r)}(\boldsymbol{r}) - \frac{1}{\mathcal{H}}\partial_r (\boldsymbol{v}\cdot \boldsymbol{n}) + A_{\alpha} (\boldsymbol{v}\cdot\boldsymbol{n}),\label{eq:delta_doppler}
\end{equation}where
\begin{equation}
    A_{\alpha} = \frac{5s_{\alpha} - 2}{\mathcal{H} r} + b_e -  \frac{\mathcal{H}'}{\mathcal{H}^2} - 5s_{\alpha},
\end{equation}is called the Doppler term, $\mathcal{H}^{-1}\partial_r (\boldsymbol{v}\cdot \boldsymbol{n})$ is the standard Kaiser term, $\mathcal{H} = aH$ is the comoving Hubble factor, $s_{\alpha} \propto \partial_{\ln r} \ln(r^2 \phi_{\alpha})$ is called the magnification bias and depends on the flux threshold of the survey through the selection function $\phi_{\alpha} = \phi_{\alpha}(L)$,
\begin{equation}
    b_e = -(1+z) \frac{\partial \ln \bar{n}}{\partial z},\label{eq:evolbias}
\end{equation}is the evolution bias and $b_{\alpha}$ is the linear bias. With the exception of the true density perturbation $\delta_{\alpha}$, all other terms appear due to departures from a perfect FLRW universe. 

Within the linear theory, we can relate quantities in configuration space with their Fourier counterpart to arrive at the main equation
\begin{eqnarray}
    \delta_{\alpha}^{(s)}(\boldsymbol{k}) = \delta^{(r)}(\boldsymbol{k})\left[ b_{\alpha} +  f \mu_{\boldsymbol{k}}^2 + i f (\mathcal{H}k^{-1}) A_{\alpha} \mu_{\boldsymbol{k}}\right],\label{eq:delta_k}
\end{eqnarray}with $\mu_{\boldsymbol{k}} \equiv (\hat{\boldsymbol{k}}\cdot\hat{\boldsymbol{r}})$ to keep the explicit dependence with the LOS. 

Assuming that all objects in the survey possess the same LOS, i.e. $\hat{\boldsymbol{k}}\cdot\hat{\boldsymbol{r}} = \mu$ is a constant (flat-sky approximation), the cross-spectrum $P_{\alpha\beta}^{(s)}(\boldsymbol{k}) = \langle \delta_{\alpha}(\boldsymbol{k}) \delta^*_{\beta}(\boldsymbol{k}) \rangle$ of two tracers $\alpha$ and $\beta$ is given by
\begin{align}
    P_{\alpha\beta}^{(s)}(\boldsymbol{k}) &= P^{(r)}(k)\Big\lbrace(b_\alpha + f\mu^2)(b_{\beta}+f\mu^2) + A_{\alpha}A_{\beta} f^2 \mu^2 \frac{\mathcal{H}^2}{k^2} \nonumber\\
    &{}\hspace{5mm} + if\mu \left[(b_{\beta}+f\mu^2)A_{\alpha} - (b_{\alpha}+f\mu^2) A_{\beta}\right]\frac{\mathcal{H}}{k}\Big\rbrace.
\end{align}In this equation, $\alpha$ and $\beta$ refers to distinct tracers, which could be different types of galaxies or dark matter halos of different masses, $f$ is the growth rate, parametrised by $f(z) \sim \Omega_m(z)^{\gamma}$, with $\gamma$ being the growth index, and $P^{(r)}(k)$ is the matter power spectrum in real space.

In this case, isotropy is broken by the choice of LOS and we can expand $P_{\alpha\beta}^{(s)}(\boldsymbol{k}) = P^{(s)}_{\alpha\beta}(k,\mu)$ in a Legendre series:
\begin{equation}
    P^{(s)}(k,\mu) = \sum_{\ell = 0}^{\infty} P_{\ell}^{(s)}(k)\mathcal{L}_{\ell}(\mu),\label{eq:legendre}
\end{equation}where
\begin{equation}
    P_{\ell}^{(s)}(k) \equiv P^{(r)}(k) \,\, c_{\ell}.\label{eq:power_multipole_coeff}
\end{equation}

\begin{figure}
    \centering
    \includegraphics[width=\columnwidth]{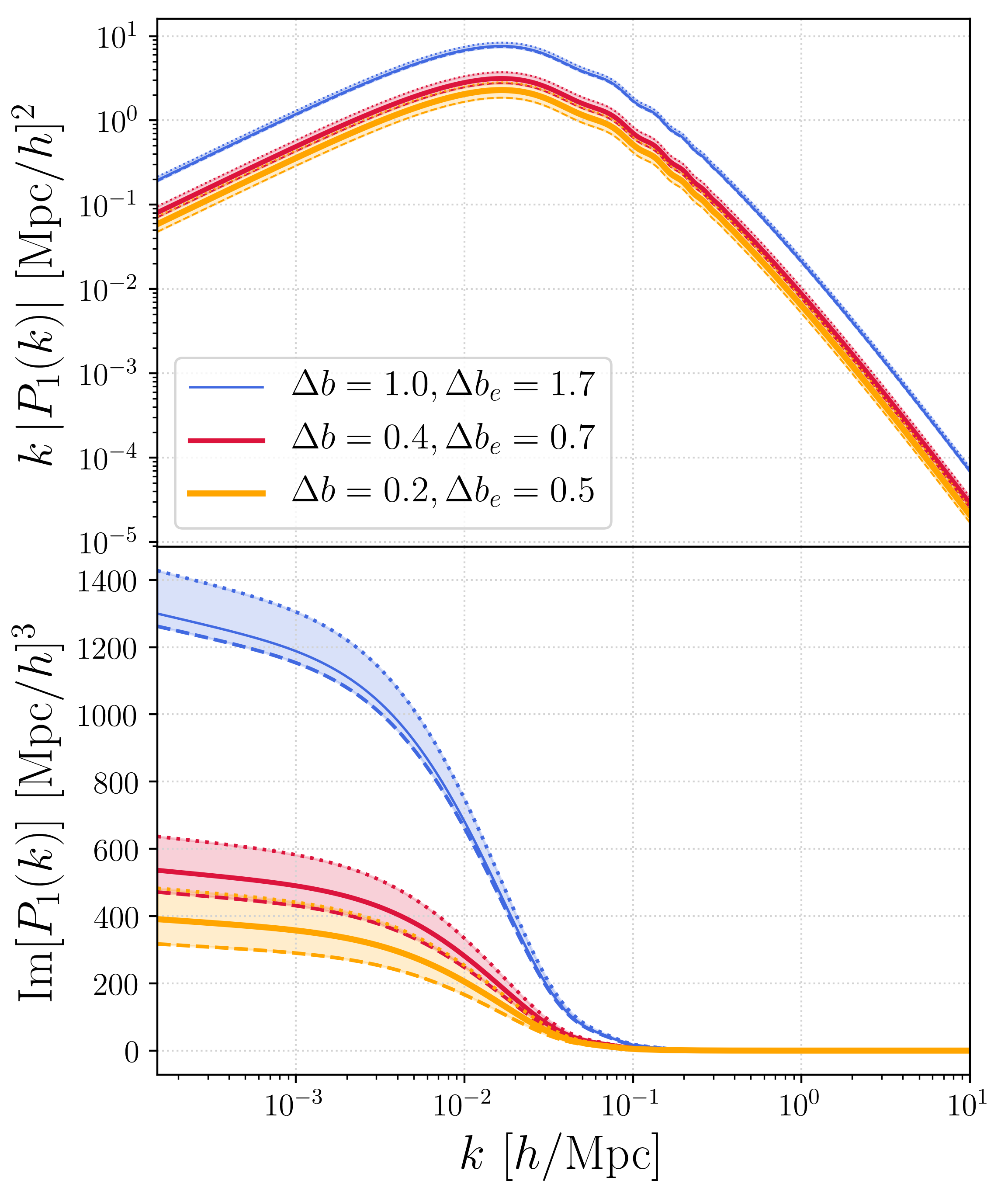}
    \caption{Theoretical prediction for the cross power spectrum dipole of different tracers at redshift $z=1.9$, with the difference in linear and evolution bias shown in the legend. Solid lines represent the case where there is no magnification bias $s_{\alpha} = 0$, whereas shaded regions represent the effect of different magnification biases among the tracers. Dotted lines show the limiting case where $s_{\alpha}$ is smaller than $s_{\beta}$ by 40\%, whilst the dashed ones show the opposite case, with $s_{\alpha}$ larger than $s_{\beta}$ by a factor of 40\%.}
    \label{fig:theory_dipole_mag}
\end{figure}

\begin{figure}
    \centering
    \includegraphics[width=\columnwidth]{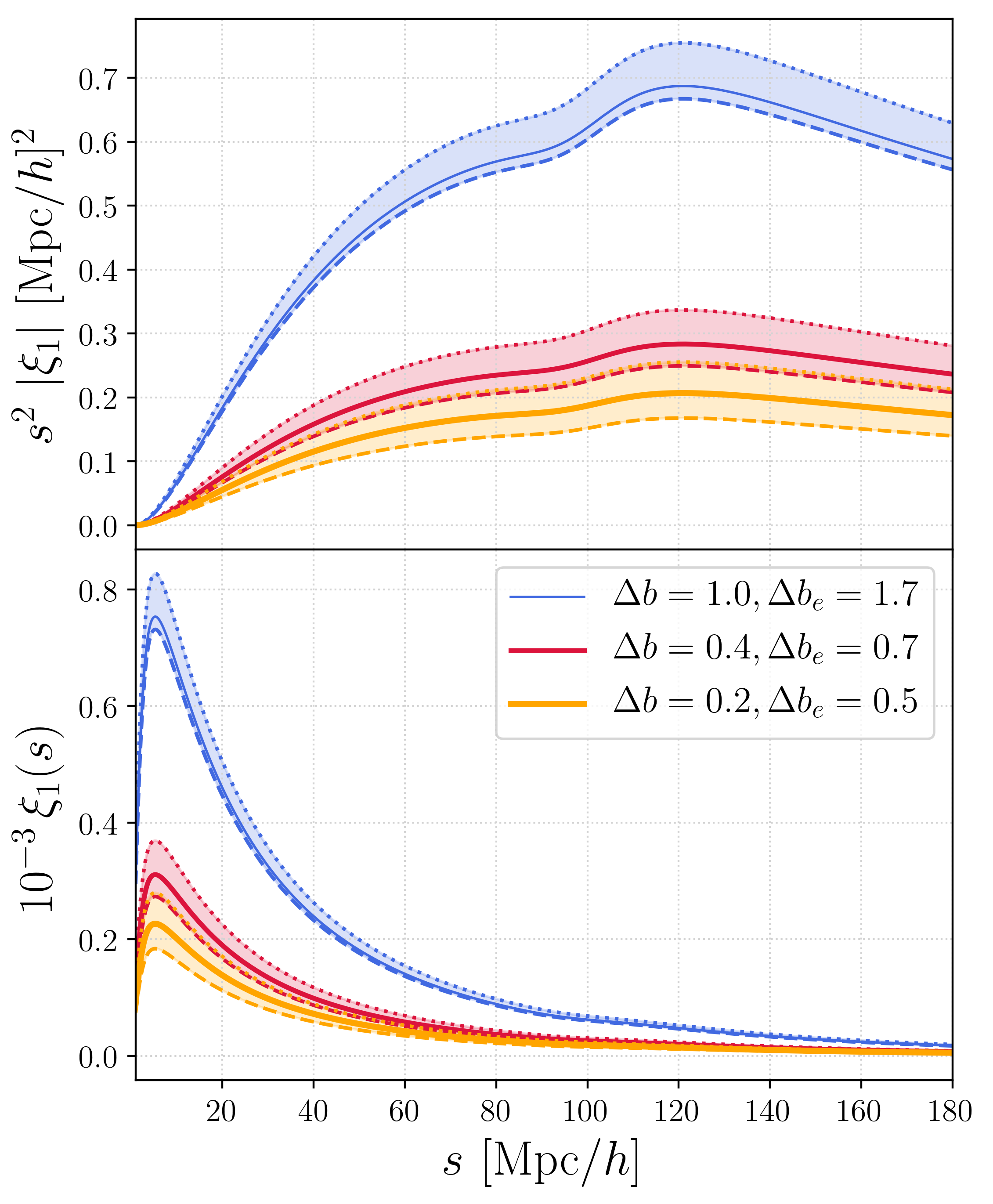}
    \caption{Same as Figure \ref{fig:theory_dipole_mag}, but for the cross-correlation function dipole of different tracers at redshift $z=1.9$. Differences in the linear and evolution bias are shown in the legend. Solid lines represent the case where there is no magnification bias $s_{\alpha} = 0$, whereas shaded regions represent the effect of different magnification biases among the tracers. Dotted lines show the limiting case where $s_{\alpha}$ is smaller than $s_{\beta}$ by 40\%, whilst the dashed ones show the opposite case, with $s_{\alpha}$ larger than $s_{\beta}$ by a factor of 40\%.}
    \label{fig:multipoles_corrfun}
\end{figure}

Neglecting the quadratic terms $\mathcal{O}(\mathcal{H}/k)^2$, the coefficients of the expansion are given by\footnote{These second order effects have a contribution smaller than 0.03\% at the largest scales probed in this work. Therefore, they shall not be considered.}
\begin{align}
    c_0(f,b) &= b_{\alpha} b_{\beta} + \frac{1}{3} f(b_{\alpha} + b_{\beta}) + \frac{1}{5} f^2,\label{eq:kaiser_mono}\\
    c_1(k,f,b,A) &=  \frac{1}{5}if\frac{\mathcal{H}}{k}\left[A_{\alpha}(3f+5b_{\beta}) - A_{\beta}(3f+5b_{\alpha})\right],\\
    c_2(f,b) &= \frac{2}{3}f(b_{\alpha}+b_{\beta}) + \frac{4}{7}f^2,\\
    c_3(k,f,A) &= \frac{2}{5}if^2\frac{\mathcal{H}}{k}(A_{\alpha}-A_{\beta}),\\
    c_4(f) &= \frac{8}{35}f^2.
\end{align}
In the absence of these quadratic corrections, the monopole, quadrupole and hexadecapole are the same as in the Newtonian case. Still, the imaginary term appearing from the relativistic corrections in equation (\ref{eq:delta_doppler}) gives rise to the dipole term manifested in the cross-spectrum of LSS tracers:
\begin{equation}
    P_1^{\alpha\beta}(k) = i \frac{f}{5}\frac{\mathcal{H}}{k}\left[A_{\alpha} (3f+5b_{\beta}) - A_{\beta} (3f+5b_{\alpha})\right] P^{(r)}(k).\label{eq:crossdipole}
\end{equation}
While it scales as $\mathcal{H}/k$ for the cross-correlation of LSS tracers, a fact that makes this signal a smoking gun for relativistic effects in the galaxy clustering, it is identically zero for the autocorrelation. We also call the reader's attention to the fact that this dipole term is antisymmetric, meaning that $\langle \delta_{\alpha}(\boldsymbol{k})\delta_{\beta}^*(\boldsymbol{k})\rangle = -\langle \delta_{\beta}(\boldsymbol{k})\delta_{\alpha}^*(\boldsymbol{k})\rangle$. In Figures \ref{fig:theory_dipole_mag} and \ref{fig:multipoles_corrfun} we illustrate the dipole term in both the Fourier and configuration spaces, respectively, for three linear and evolution bias differences (different colours) at a fixed redshift \CG{of $z=1.9$}.

In what follows we explore the detection \CG{of the signal given by equation} (\ref{eq:crossdipole}) in a relativistic simulation of a light cone, described in Section \ref{sec:sim}. Since we will be dealing with dark matter halos, the magnification bias $s_{\alpha}$ in the Doppler term vanishes, as there is not flux limit in our samples. Therefore, in addition to the linear bias of the halos, the remaining parameter entering the theoretical predictions is the evolution bias, \CG{equation} (\ref{eq:evolbias}). The procedure for fitting $b_e$ from the halo samples is described in Appendix \ref{ap:evolbias}, with the results discussed in Section \ref{sec:ps}.

\section{Simulation}\label{sec:sim}

In this work we make use of a large $N$-body simulation performed with the relativistic code \textit{gevolution} \citep{adamek2016a, adamek2016b}. The simulation has a comoving volume of (2.4 Gpc$/h)^3$ with dark matter particles of mass $2.64\times 10^9 \,(M_{\odot}/h)$, and represents a typical $\Lambda$CDM cosmology: $h = 0.67556$, $\omega_b = 0.022032$, $\omega_{\mathrm{cdm}} = 0.12038$, $T_{\mathrm{CMB}} = 2.7255$ K, $A_s= 2.215 \times 10^{-9}$, $n_s = 0.9619$, $N_{\mathrm{ur}} = 3.046$, and $N_{\mathrm{ncdm}} = 0$. In order to avoid replications in the light cone, the pencil beam was carefully oriented in the periodic domain. The initial conditions for the simulation were set at a redshift of $z=127$. 

Unlike the standard approach to building light cones \citep{merson2012,smith2017,breton2019}, which consists of generating many simulation snapshots with a sufficient small redshift step between them to avoid time \CG{discretisation} effects in the final light cone, the light cone output from \textit{gevolution} records particle positions and velocities on the fly. During the simulation, particles are identified that are within a proper comoving distance interval from a pre-defined observer that would cause them to be placed in the final catalogue. These particles are then shifted by a fractional time-step and recorded on the null FLRW hypersurface given by the past light cone of the observer. Hence, there are no time \CG{discretisation} artefacts and no need to generate an enormous amount of snapshots to build the light cone. In our case, no replications whatsoever were performed in order to cover the whole light cone volume, which has the  advantage of removing any concerns about spurious correlations on large scales due to periodicity for example.

The \textit{gevolution} code does not employ the adaptive mesh refinement (AMR) method and thus has a low accuracy at small scales. However, while AMR can improve the one-halo term by better resolving halo substructures, it does not significantly impact the large scales dominated by the two-halo term, which is the focus of this work. As will be pointed out in Section \ref{sec:catalogue}, all subhalos are discarded in our analysis in any case.

\subsection{Ray tracing}\label{sec:raytracer}

We apply a ray tracing algorithm to our simulation as a post-processing tool. The algorithm was previously described in \citet{lepori2020}, but we give a brief review of it here.

The purpose of the ray tracer is to add extra information on source objects within the simulation to the catalogue, such as their angular diameter distance ($D_A$) relative to a specific observer, the respective observed redshift ($z$), or the ellipticity ($\epsilon$) which is closely related to the weak-lensing shear ($\gamma$). In contrast to the more common case where ray tracing is applied to Newtonian $N$-body simulations, in \textit{gevolution} the metric perturbations and the source positions are both provided in Poisson gauge, which makes the treatment of gauge issues transparent. Our algorithm also does not rely on the Born approximation to model the light path. Importantly, incorrectly modelling the lensing probability distribution function can lead to errors in estimating cosmological parameters, as shown in \citet{adamek2019}.

The algorithm is similar to the one presented in \citet{breton2019} and works by integrating the geodesic equations backwards in time from the observer to the source of interest on the observer's past light cone. A physical definition of source, such as a halo or a dark matter particle, is required, as a four-velocity vector is needed to define the source's rest frame. This allows us to get the observed redshift of the source in a gauge-independent way. For each of these sources, we use the background FLRW model to give us the initial direction vector ($\boldsymbol{n}$) for each light ray towards a source. We then integrate backwards in time with the fully perturbed metric until the light ray reaches its closest approach to the event on the light cone. At this point, we can now calculate a ``deflection angle'' by which the initial $\boldsymbol{n}$ must be corrected to achieve a closer approach to the source. We repeat this process several times until suitable convergence is achieved.

This process works well in the weak-lensing regime, as only a single null ray exists between the observer and each source. In the strong lensing regime, multiple images can be formed, which complicates matters. The number of sources where this phenomenon is observed is negligible however, and so we concentrate only on weak lensing. Since strong lensing will only affect our results on very small scales where an image could be duplicated, this choice has a negligible impact on our analysis.

Ray tracing is the key step in properly incorporating relativistic corrections in our analysis. For example, instead of using the redshift output directly from the halo finder which would only include the background expansion and the Doppler correction, we are able to use the `observed' redshift, which includes all relativistic effects. We can also calculate the perturbed position of sources on the sky, which is important for any \CG{$n$}-point correlation calculations done using the catalogue. The algorithm also output $D_A$ and both the real and the imaginary parts of the shear component separately ($\gamma_{1}+\mathrm{i} \gamma_{2} \simeq-\frac{\epsilon}{4}$), although these are not needed in the current analysis.

\subsection{Halo catalogue}\label{sec:catalogue}
From the real space particles, the halo catalogue was created with the Rockstar halo finder \citep*{rockstar}, using a friends-of-friends (FOF) algorithm with linking length $b=0.28$ in order to detect over $10^7$ halos in the light cone. 

After going through the ray-tracer algorithm, which is crucial to connect the halos and the observer, the perturbed three-dimensional positions of halos were obtained and the resulting file consists of three mock surveys contained within the range $0.0 \lesssim z \lesssim 7.1$, with different survey areas. The survey that will be used in this work spans the range of comoving look-back distance from 275 up to 4560 $\mathrm{Mpc}/h$. 

We limit ourselves to the high-redshift region between $z_{\mathrm{min}} = 1.7$ and $z_{\mathrm{max}} = 2.9$, with redshift bins of size $\Delta z = 0.4$ \CG{that} kept the variation of the growth function within the 5\% limit\footnote{This criterion was chosen to keep halos of different evolutionary stages somewhat separated.}. Each redshift bin has an effective volume of $\sim 0.7$ (Gpc/$h$)$^3$ given the chosen cosmology and the sky fraction $f_{\mathrm{sky}} \sim 0.01$. After this redshift selection, we were left with $8.5\times 10^6$ dark matter halos.

The high-redshift binning was chosen to deliver a reasonable volume necessary for the observation of the relativistic features at large scales, giving an effective fundamental mode of $k_{\mathrm{F}} = 2\pi/{V^{1/3}} \sim 7 \times 10^{-3}\, (h/\mathrm{Mpc})$. In a future work we will present the results of the same analysis, but in the full-sky case. The current survey area of $\sim 400$ deg$^2$ is compatible with the current survey areas available for a cross-correlation analysis \citep{zhao2020}.

The final halo catalogue was then separated into three halo samples per redshift bin, \CG{each of them with different masses} such that, at each redshift bin, the number of halos was the same \CG{for} each sample. The main properties of these samples are detailed in Table \ref{tab:data_eqn_hz_dz04}. Because more massive halos are expected at lower redshifts, the effective redshift $\bar{z}$ of each halo sample varies slightly, but only by less than 0.5\%; therefore, we considered the values shown in the Table as the respective central redshift. The biases were computed by fitting the ratio between the real\CG{-}space power spectrum of the halos and dark matter [see Appendix \ref{ap:halos} for a throughout discussion and comparison with the \citet{tinker2010} fitting function]. The halo population incorporating all halos is referred to as $H_{\mathrm{all}}$ in what follows.

\begin{table}
    \small
    \centering
    \caption{Specifications of the halo samples, selected to match the number density for each population, yielding $\bar{n}_0 \sim \bar{n}_1 \sim \bar{n}_2$. The mean redshift $\bar{z}$ is obtained from all the halos within each redshift bin, as the effective redshift of each halo sample differs from $\bar{z}$ in a sub-per cent level ($\lesssim 0.5\%$). The biases have been computed from the monopole of the power spectrum by fitting a linear polynomial to the ratio between the halo auto-spectra and the real-space linear matter power spectrum (see Appendix \ref{ap:bias}), differing from the Tinker bias by $\sim 5\%$. For the three redshift bins, the volumes are such that the fundamental mode of observation is $k_{\mathrm{F}} = 2\pi/{V^{1/3}} \sim 7 \times 10^{-3}\, (h/\mathrm{Mpc})$.}
    \begin{tabular}{l c c c c}
        \hline
             & \# &  Mean mass &  Bias & $\bar{n}(\bar{z})$ \\
         & halos & $[M_{\odot}/h]$ & (fit) & $[\mathrm{Mpc}/h]^{-3}$ \\
        \hline
        \multicolumn{5}{c}{$\bar{z} = 1.89$} \\
        \hline
        All & 480643 &  $4.41 \times 10^{12}$ & 2.927 & 7.053 $\times10^{-4}$\\
        H$_{0}$  & 160081 &  $1.86\times 10^{12}$ & 2.551 & 2.349 $\times10^{-4}$ \\
        H$_{1}$  & 160547 &  $2.83\times 10^{12}$ & 2.758 & 2.356 $\times 10^{-4}$\\
        H$_{2}$  & 160015 &  $8.54\times 10^{12}$ & 3.477 & 2.348 $\times 10^{-4}$ \\
        \hline
        \multicolumn{5}{c}{$\bar{z} = 2.29$} \\
        \hline
        All & 326899 &  $3.85 \times 10^{12}$ & 3.469 & 4.666 $\times10^{-4}$\\
        H$_{0}$  & 109003 &  $1.83\times 10^{12}$ & 3.020 & 1.556 $\times10^{-4}$ \\
        H$_{1}$  & 108809 &  $2.66\times 10^{12}$ & 3.270 & 1.553 $\times 10^{-4}$ \\
        H$_{2}$  & 109087 &  $7.05\times 10^{12}$ & 4.154 & 1.557 $\times 10^{-4}$ \\
        \hline
        \multicolumn{5}{c}{$\bar{z} = 2.69$} \\
        \hline
        All & 205678 &  $3.44 \times 10^{12}$ & 4.214 & 2.947 $\times10^{-4}$\\
        H$_{0}$  & 68501 &  $1.80\times 10^{12}$ & 3.735 & 9.815 $\times10^{-5}$\\
        H$_{1}$  & 68550 &  $2.52\times 10^{12}$ & 4.006 & 9.822 $\times 10^{-5}$ \\
        H$_{2}$  & 68627 &  $5.98\times 10^{12}$ & 4.932 & 9.833 $\times 10^{-5}$ \\
        \hline
    \end{tabular}
    \label{tab:data_eqn_hz_dz04}
\end{table}

\section{Power spectrum multipole estimator}\label{sec:ps}

To compute the power spectrum multipoles we make use of the standard approach proposed by \citet{yamamoto2006, bianchi2015} and \citet{scoccimarro2015} \citep*[for pioneering work see also][]{yamamoto2000}, which we dub YBS estimator. It is built upon the practical algorithm developed by \citet*{FKP} to optimally estimate the power spectrum of galaxy surveys with a varying selection function. As mentioned in Section \ref{sec:theory}, the selection function $\bar{N}$ encodes the spatial modulations of the mean number density of objects. For both spectroscopic and photometric surveys, the selection function accounts for all non-cosmological effects, being sensitive, for example, to the different intrinsic brightness of galaxies. 

The selection function gives an estimate of the probability that a galaxy brighter than a certain threshold, at a distance $\boldsymbol{s}$, is included in the sample. Hence, it is intrinsically related to the notion of luminosity function $\Phi(L)$ \citep{martinez}. In \citet*{wang2020} a clear example of such fact is given, with the luminosity function of eBOSS quasars (QSO) being used to fit the QSO number density and derive the evolution and magnification biases.

To resume the construction of the estimator, $N_{X}(\boldsymbol{x}_{ijk})$ denotes either the count-in-cells of \CG{the random, $X=r$, or of} the data (halo) catalogue, $X=h$, where $\boldsymbol{x}_{ijk}$ is the position of each cell in a three-dimensional grid obtained by a mass assignment scheme, e.g. Nearest Grid Point (NGP), Cloud In Cell (CIC), or Triangular Shaped Cloud (TSC). In this analysis we consider the simplest NGP assignment.

We begin by defining the weighted galaxy fluctuation, or the overdensity field\footnote{It is more instructive to write $F(\boldsymbol{x}) = w(\boldsymbol{x})[n(\boldsymbol{x}) - \bar{n}(\boldsymbol{x})] = w(\boldsymbol{x})\bar{n}(\boldsymbol{x})\delta(\boldsymbol{x}) = W(\boldsymbol{x}) \delta(\boldsymbol{x})$, where we call $W(\boldsymbol{x})$ the window function. Then, in Fourier space $F(\boldsymbol{k})$ is the convolution of the window with the density contrast: $F(\boldsymbol{k}) = (2\pi)^{-3} \int \mathrm{d}^3 q\, W(\boldsymbol{k}-\boldsymbol{q})\delta(\boldsymbol{q})$, and one can show that $\langle F(\boldsymbol{k}) F(-\boldsymbol{k})\rangle = (2\pi)^{-3} \int \mathrm{d}^3 q\, |W(\boldsymbol{k}-\boldsymbol{q})|^2 P(\boldsymbol{q}) + \int \mathrm{d}^3 x\, w^2(\boldsymbol{x})\bar{n}(\boldsymbol{x})$. Therefore, in \citet{FKP} it is considered the overdensity field divided by the magnitude of the window function, $W^2 \equiv \int \mathrm{d}^3 x\, W^2(\boldsymbol{x})$, which we called $\mathcal{N}$ \citep{jeong2010}.}, as
\begin{equation}
    F(\boldsymbol{x}) = \frac{w(\boldsymbol{x})}{\mathcal{N}} \left[n_h(\boldsymbol{x}) - \alpha n_r(\boldsymbol{x})\right],
\end{equation}where $n_h(\boldsymbol{x}) = \sum_{i=1}^{N_h}\delta^D(\boldsymbol{x}-\boldsymbol{x}_i)$ is the number density \CG{that} will be written as a grid, after \CG{a} mass assignment scheme is chosen. Therefore, in practice $n_h(\boldsymbol{x}) = N_h(\boldsymbol{x}_{ijk})$ is the count-in-cells grid and $n_r(\boldsymbol{x})$ is the corresponding quantity for the random catalogue, which is obtained by randomly sampling $\alpha^{-1}$ times more objects within the survey volume, with the same selection function as the real data. 

The results presented here do not employ a weighting scheme, i.e. $w(\boldsymbol{x}) = 1$, and we follow \citet{jeong2010} for the implementation of the quadratic estimator. The normalisation factor will be given by 
\begin{equation}
    \mathcal{N} \approx \frac{\alpha^2}{\ell_x \ell_y \ell_z} \sum_{\boldsymbol{x}_{ijk}} N_r^2(\boldsymbol{x}_{ijk}),
\end{equation}and the shot noise, only relevant for the monopole term, will be 
\begin{equation}
    P_{\mathrm{shot}} \approx \ell_x \ell_y \ell_z\left(\frac{1+\alpha}{\alpha}\right)\frac{\sum_{\boldsymbol{x}_{ijk}} N_r(\boldsymbol{x}_{ijk})}{\sum_{\boldsymbol{x}_{ijk}} N_r^2(\boldsymbol{x}_{ijk})},
\end{equation}where $\ell_i \equiv L_i/n_i$ is the size of each cell dimension in units of Mpc$/h$. In this work we choose $\ell_i = 10$ Mpc$/h$.

\begin{figure}
    \centering
    \includegraphics[width=\columnwidth]{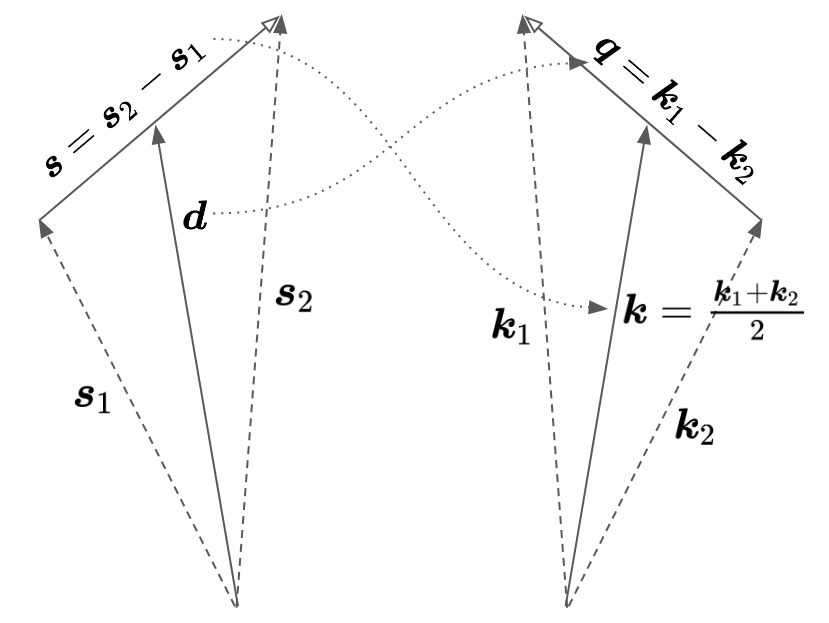}
    \caption{The mirrored scheme at the right\CG{-}hand side corresponds to the Fourier transform of the left\CG{-}hand configuration. The Fourier counterpart of $\boldsymbol{d}$ and $\boldsymbol{s}$ are, respectively, $\boldsymbol{q}$ and $\boldsymbol{k}$ \citep*{reimberg2016}. On the left side\CG{,} the configuration for the local estimator $P(\boldsymbol{k},\boldsymbol{d})$ is depicted. The YBS estimator corresponds to integrating over all possible lines-of-sight $\boldsymbol{d}$ and averaging over $k$-bins. In this sketch, $\boldsymbol{s} = \boldsymbol{s}_2 - \boldsymbol{s}_1$, with the observer located at the lower vertex.}
    \label{fig:config}
\end{figure}

After the construction of these quantities, the power spectrum multipoles can be obtained by the YBS estimator, which we now briefly discuss. The whole idea of this method relies in \CG{generalising} the power spectrum to \textit{local} regions in space, where statistical homogeneity may be assumed. These regions are defined by a single middle line-of-sight $\boldsymbol{d} = (\boldsymbol{s}_1 + \boldsymbol{s}_2)/2$, as shown on the left of Figure \ref{fig:config}. Then, the corresponding power spectrum at this region is
\begin{equation}
    P(\boldsymbol{k}_1,\boldsymbol{k}_2) = \int \mathrm{d}^3s_1 \int \mathrm{d}^3s_2 \,\,\xi(\boldsymbol{s}_1,\boldsymbol{s}_2)\, \mathrm{e}^{i\boldsymbol{k}_1\cdot\boldsymbol{s}_1}\mathrm{e}^{-i\boldsymbol{k}_{2}\cdot\boldsymbol{s}_2}.
\end{equation}Notice that
\begin{equation}
    \mathrm{e}^{i\boldsymbol{k}_1\cdot\boldsymbol{s}_1}\mathrm{e}^{-i\boldsymbol{k}_{2}\cdot\boldsymbol{s}_2} = \mathrm{e}^{i\boldsymbol{d}\cdot \boldsymbol{q}}\mathrm{e}^{-i\boldsymbol{k}\cdot\boldsymbol{s}},
\end{equation}where the Fourier transform of the local configuration is shown on the right of Figure \ref{fig:config}. With this change of coordinates, one can see that the local power spectrum can be obtained by taking the Fourier transform of the first component of the local correlation function $\xi(\boldsymbol{s}_1,\boldsymbol{s}_2) = \xi(\boldsymbol{s},\boldsymbol{d})$. Finally, the multipoles of the local power spectrum can be obtained from the Legendre expansion
\begin{equation}
    P(\boldsymbol{k},\boldsymbol{d}) = \sum_{\ell} P_{\ell}(k,d) \mathcal{L}_{\ell}(\hat{\boldsymbol{k}}\cdot\hat{\boldsymbol{d}}).
\end{equation}Lastly, the inversion of this relation yields the power spectrum multipoles:
\begin{align}
    \hat{P}_{\ell}(k) &= \Big\langle\frac{2\ell + 1}{2\mathcal{N}} \int  \mathrm{d}^3 s_1 \int \mathrm{d}^3 s_2 \, F(\boldsymbol{s}_1) F(\boldsymbol{s}_2) \nonumber\\
    &{}\hspace{15mm} \times \mathrm{e}^{-i\boldsymbol{k}\cdot(\boldsymbol{s}_2 - \boldsymbol{s}_1)} \mathcal{P}_{\ell}(\hat{\boldsymbol{k}}\cdot\hat{\boldsymbol{d}}) - S_{\ell} (\boldsymbol{k})\Big\rangle,
\end{align}where the brackets correspond to an average over $k$-shells and $S_{\ell}$ is the shot-noise term, only relevant for the monopole.

In order to speed up the computation of the multipoles by means of FFTs, the YBS estimator takes the end-point LOS $\boldsymbol{s}_1$. It is worth mentioning that this LOS intrinsically generates odd multipoles that may impact the signal we are trying to measure. Hence, this must be accounted for in the window function, as discussed in Appendix \ref{ap:window}.

With the adoption of this LOS, the monopole can estimated as
\begin{equation}
    \hat{P}_0(k) = \frac{1}{\mathcal{N}} \langle F_0(\boldsymbol{k}) F_0^{*}(\boldsymbol{k}) - S_0 \rangle,
\end{equation}where
\begin{equation}
    F_0(\boldsymbol{k}) = \int \mathrm{d}^3x \, F(\boldsymbol{k}) \,\mathrm{e}^{i\boldsymbol{k}\cdot\boldsymbol{x}}
\end{equation}is the Fourier transform of the overdensity field with no weight,
\begin{equation}
    F(\boldsymbol{x}) = n_h(\boldsymbol{x}) - \alpha n_r(\boldsymbol{x}),
\end{equation}and the dipole is obtained by
\begin{equation}
    \hat{P}_1(k) = \frac{3}{\mathcal{N}} \langle F_0(\boldsymbol{k}) F_1^{*}(\boldsymbol{k})\rangle,
\end{equation}with
\begin{equation}
    F_1(\boldsymbol{k}) = \sum_{i = x,y,z} \hat{k}_i f_{1,i}(\boldsymbol{k}),
\end{equation}and 
\begin{equation}
    f_{1,i}(\boldsymbol{k}) = \int \mathrm{d}^3 r \, \hat{r}_i F(\boldsymbol{r})\,\mathrm{e}^{i\boldsymbol{k}\cdot\boldsymbol{r}}.
\end{equation}For higher order multipoles, we refer the interested reader to \citet{bianchi2015} and \citet*{beutler2019} for the even and odd ones, respectively. 

Finally, to compute the cross-dipole, $F_0(\boldsymbol{k})$ and $F_1(\boldsymbol{k})$ are built from the first and second tracers, respectively, and the normalisation becomes \citep{beutler2020}:
\begin{equation}
    \mathcal{N} \approx \frac{\alpha_{(1)}\alpha_{(2)}}{\ell_x \ell_y \ell_z} \sum_{\boldsymbol{x}_{ijk}} N_r^{(1)}(\boldsymbol{x}_{ijk})N_r^{(2)}(\boldsymbol{x}_{ijk}).
\end{equation}

\section{Results}\label{sec:results}

For the objects under analysis (dark matter halos), the concept of luminosity function can replaced by the halo mass function\footnote{See Appendix \ref{ap:halos} for a detailed discussion on the mass function of our samples.} $d\bar{n}/d\ln M$, which gives the probability of having a mean number of halos, within some comoving volume, with mass in the range $[\ln M_i,\ln M_i +\mathrm{d}M]$. Thereby, the selection function coincides with the comoving mean number density of halos within a certain mass bin $[M_i, M_{i+1}]$, in complete analogy to the definition of $\phi(L)$ from a luminosity function \citep[e.g. see][]{martinez}:
\begin{equation}
    \bar{n}(z,\Delta M) = \int_{M_i}^{M_{i+1}} \mathrm{d} \ln M \, \frac{\partial\bar{n}}{\partial \ln M}.
\end{equation}

Still for the specific case of dark matter halos, the magnification bias is identically zero, $s = 0$, and the only term accounting for the mass function variations is the evolution bias, which explores its dependency with time: this is related to the fact that halos can merge to form more massive structures and, thereby, their number counts are not conserved. The evolution biases $b_e^i(z)$ of each halo sample $i=\lbrace H_0,H_1,H_2, H_{\mathrm{all}} \rbrace$ considered in this work, within each redshift bin, is shown in Figure \ref{fig:evolbias_eq}. The procedure to compute $b_e^i(z)$ is described in Appendix \ref{ap:halos} and it is based on the work of \citet{beutler2020}. The values computed at the mean redshift of each $z$-bin, $b_e^i(\bar{z})$, are shown in Table \ref{tab:nbarz_fit_eq}.

\begin{figure*}
    \centering
    \includegraphics[width=\textwidth]{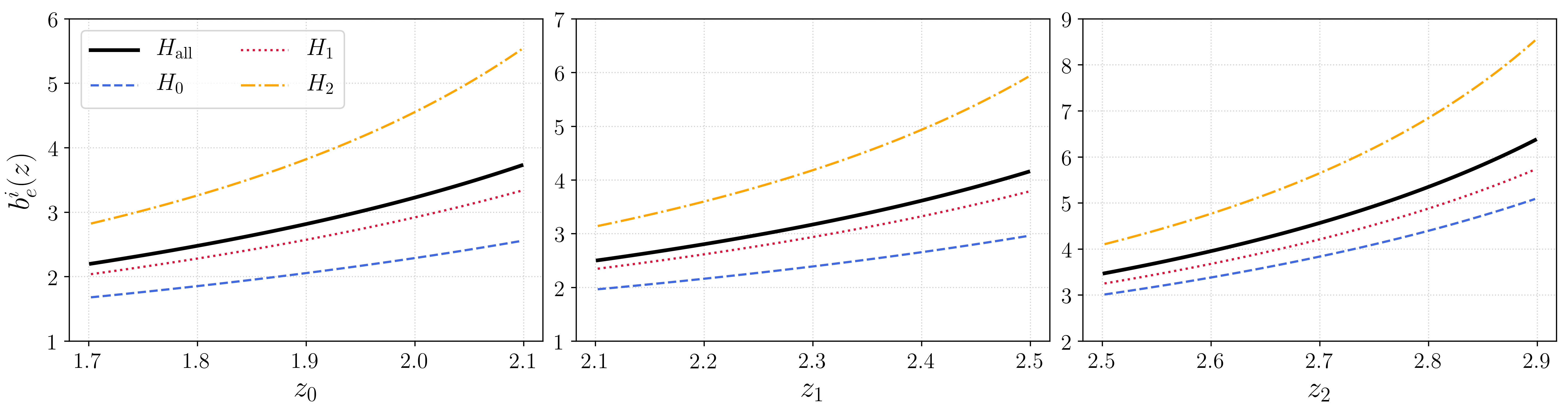}
    \caption{Evolution biases for the halo samples described in Table \ref{tab:data_eqn_hz_dz04}, at each redshift bin considered in the analysis. Notice that, even though the number density is approximately the same for all the halo populations, the intrinsic evolution of each halo population \CG{with redshift} gives rise to very different evolution biases. The computation is described in Appendix \ref{ap:evolbias}, following \citet{beutler2020}.}
    \label{fig:evolbias_eq}
\end{figure*}

We work with 14 band powers (Fourier bins) linearly spaced between $k_{\mathrm{min}} \approx 0.006$ $h/$Mpc and $k_{\mathrm{max}} \approx 0.157$ $h/$Mpc, with $\Delta k \approx 0.01\,\, h/\mathrm{Mpc} \approx 1.67\, k_{\mathrm{F}}$. As already mentioned in Section \ref{sec:ps}, we work with three-dimensional grids containing $n_x = n_y, n_z < n_x$ cells of side $\ell_i = 10$ Mpc$/h$, with the number of cells $n_i$ varying between the different redshift bins. This binning was chosen to deliver a less noisy measurement at large scales. 

\begin{figure}
    \centering
    \includegraphics[width=\columnwidth]{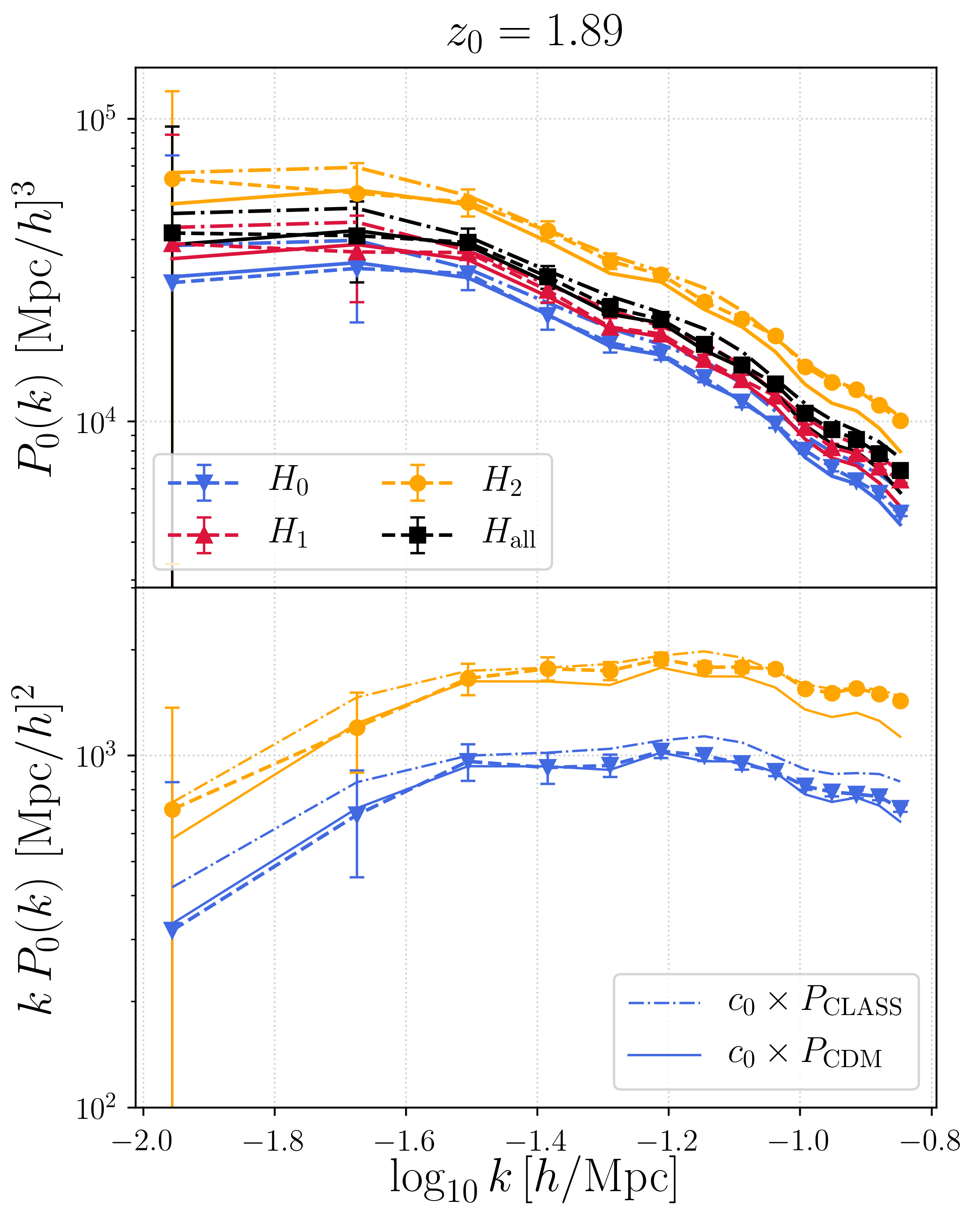}
    \caption{\CG{Monopoles} estimated from the four halo samples in the first redshift bin: $\bar{z} \approx 1.89$. Error bars are computed from the standard deviation of 100 log-normal mocks generated with the same box dimensions, evolution and linear biases, selection function\CG{,} and survey mask. \CG{Solid lines represent the redshift-space halo power spectrum computed from the estimated real-space CDM spectrum}. The bottom panel shows $kP_0(k)$ for the \CG{least} and \CG{most} massive halo samples (lower and upper \CG{curves}, respectively) for better visibility of the errors and the larger scales; it also shows the theoretical monopoles without considering the window function computed (dash-dotted \CG{curves}).}
    \label{fig:mono_z0}
\end{figure}

Figure \ref{fig:mono_z0} shows the \CG{monopoles} estimated from the four halo samples in the first redshift bin ($\bar{z} \approx 1.89$). Apart from the amplitude of the monopole, not much change occurs between different redshifts, hence we only show the first $z$-bin here. As we discuss in Appendix \ref{ap:window}, in particular as shown in Figure \ref{fig:convolution}, the window function had an impact of the order of 5\% or less at the scales we considered here. Because its impact is larger for $k \lesssim 0.01$ $h/$Mpc, where the measurement is completely dominated by cosmic variance, we find it more instructive to compare the halo monopole with the one estimated from the real-space power spectrum of cold dark matter (CDM) particles multiplied by the proper coefficient of the Legendre expansion, equation (\ref{eq:kaiser_mono}). This shows us that the theoretical connection between the CDM particles in real space is consistent with the halo measurements obtained from the redshift-space catalogue. This also has the advantage of naturally incorporating any window function effect one might be concerned with. For completeness, we also compare the measurements with the same monopole coefficient, but using the linear real-space power spectrum extracted from the CLASS Boltzmann solver \citep*{blas2011}. The bias parameters used for CLASS are presented in Table \ref{tab:data_eqn_hz_dz04} and it is used throughout this work. To compute the monopole coefficient with the real-space CDM power spectrum, we consider the bias parameters shown in Table \ref{tab:bias_eqbin}, fourth column (CDM), and refer the reader to Appendix \ref{ap:bias} for more details.

We estimate the error bars from the standard deviation of 100 log-normal mocks generated with the same characteristics of the original data: box dimensions, evolution and linear biases, selection function\CG{,} and survey mask. For this last step, we first generated the log-normal mocks for the whole box encompassing the different redshift bins of the light cone, and then applied the proper mask to select the specific angular region. However, because these error bars only quantify the variance of the estimator, the particular survey features do not matter much, and thus it should be possible to compute the variance of mocks inside the box, with the only caveat of following the mean number density of tracers to properly incorporate the shot noise. The impact of the window function in the standard deviation of the log-normal samples generated minor changes at very large scales, and we do not explore this further.

To what concerns the variance of our measurements, the window function had a negligible impact. Nonetheless, to avoid any sort of complications, we explore the asymmetry of the relativistic signal as suggested in \citet{beutler2020}: by computing $\Delta P_1 = P_1^{\alpha\beta} - P_1^{\beta\alpha}$ it is possible to isolate the relativistic contribution and get rid of the impact of the window function, which is symmetric. As pointed in Section \ref{sec:theory}, the Doppler term is antisymmetric, $\langle \delta_{\alpha}(\boldsymbol{k})\delta_{\beta}^*(\boldsymbol{k})\rangle = -\langle \delta_{\beta}(\boldsymbol{k})\delta_{\alpha}^*(\boldsymbol{k})\rangle$, and thus $\Delta P_1 \sim 2\langle \delta_{\alpha}(\boldsymbol{k})\delta_{\beta}^*(\boldsymbol{k})\rangle$.

\begin{figure}
    \centering
    \includegraphics[width=\columnwidth]{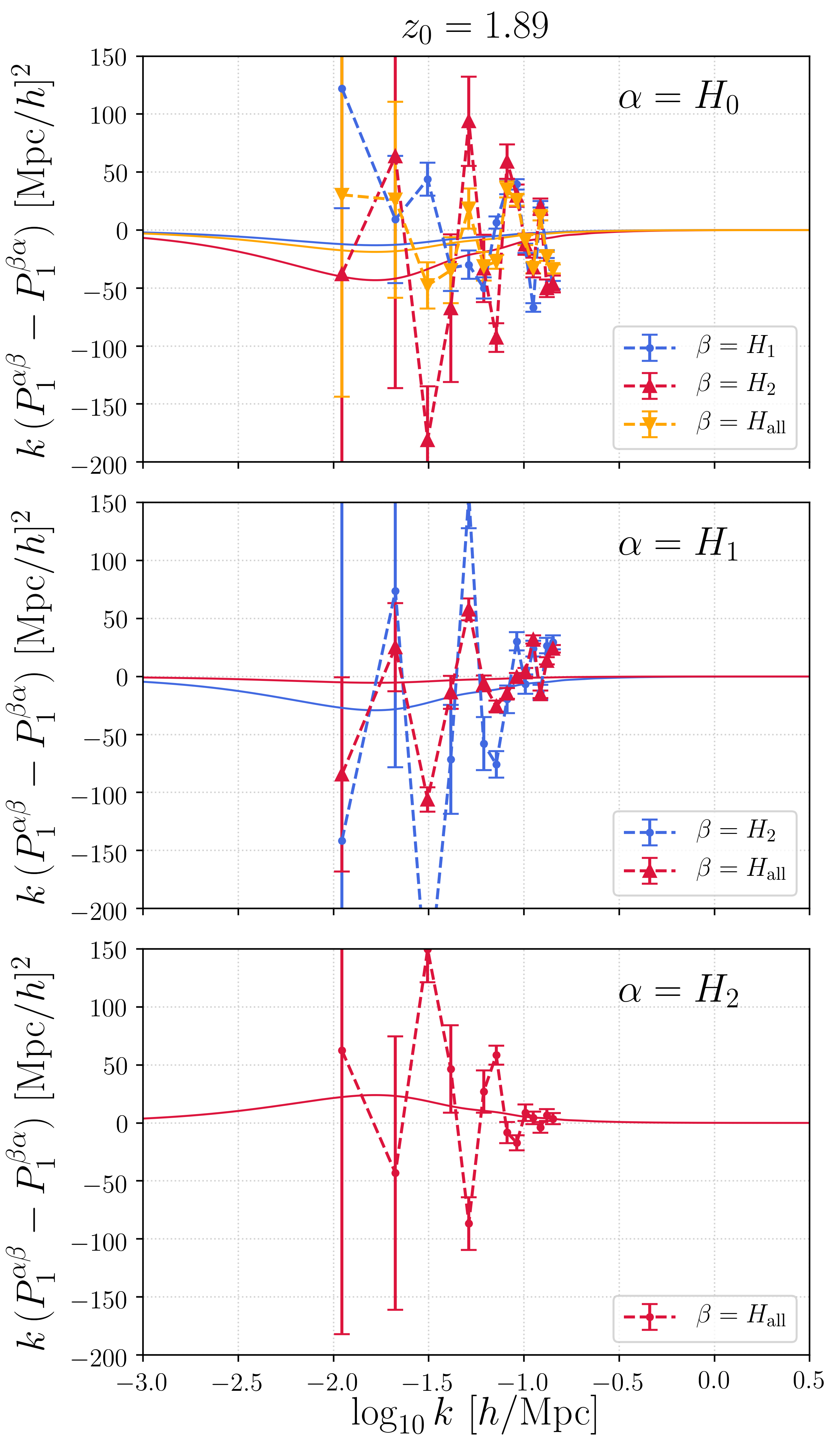}
    \caption{\CG{Dipoles} estimated from the four halo samples in the first redshift bin: $\bar{z} \approx 1.89$. Error bars are computed from the standard deviation of 100 log-normal mocks generated with the same box dimensions, evolution and linear biases, selection function and survey mask. Solid lines represent the theory, as $P_1^{\alpha\beta} - P_1^{\beta\alpha}$ is free from the window function contribution. We show all possible combinations of halos.}
    \label{fig:dipo_z0}
\end{figure}

\begin{figure*}
    \centering
    \includegraphics[width=\textwidth]{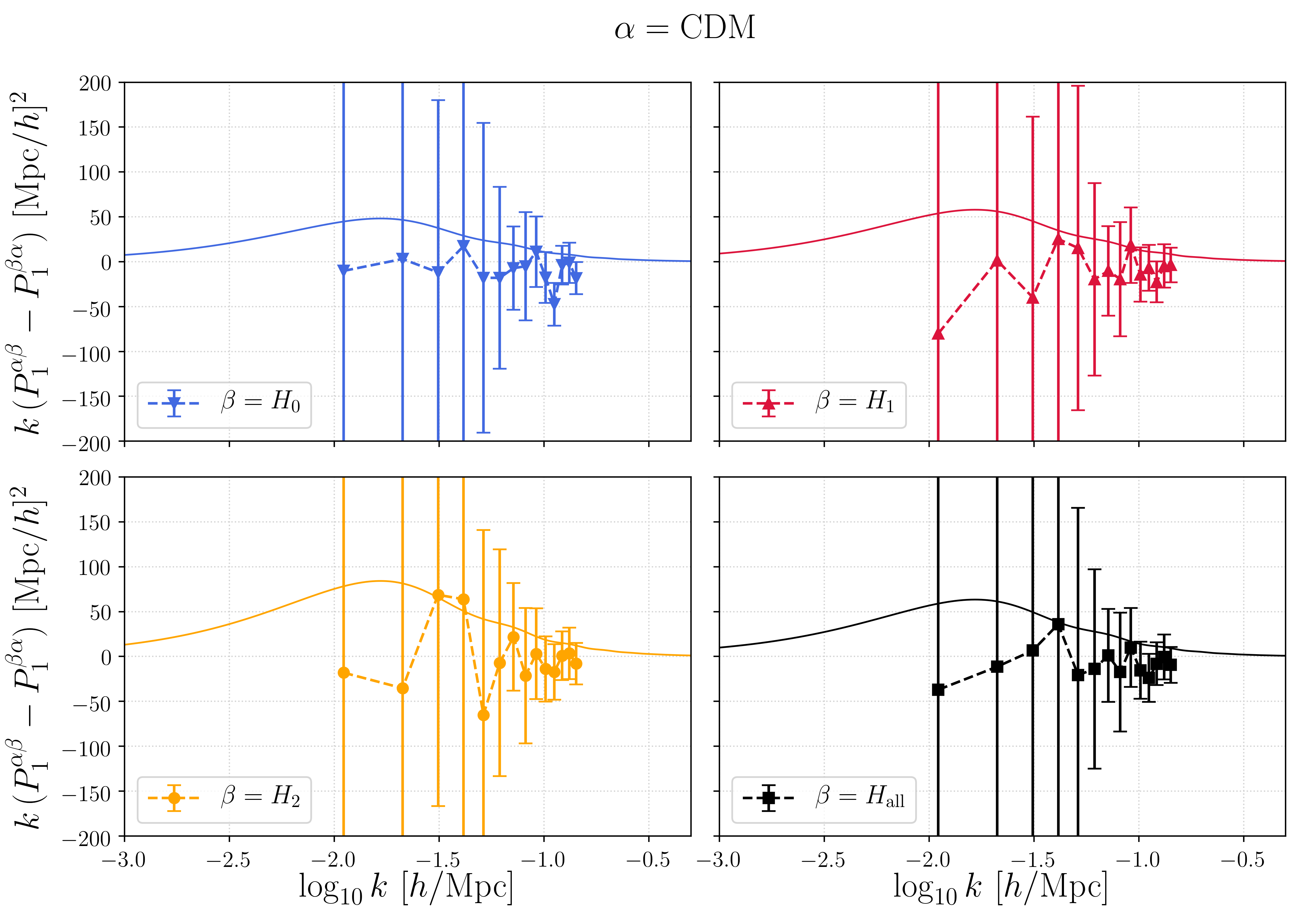}
    \caption{\CG{Dipoles} estimated from the four halo samples in the first redshift bin: $\bar{z} \approx 1.89$. Error bars are computed from the standard deviation of 100 log-normal mocks generated with the same box dimensions, evolution and linear biases, selection function and survey mask. Solid lines represent the theory, as $P_1^{\alpha\beta} - P_1^{\beta\alpha}$ is free from the window function contribution. We show all possible combinations of halos with cold dark matter.}    \label{fig:dipo_z0_cdm}
\end{figure*}

Consistently, the theoretical prediction for the last cross-correlation $H_2\times H_{\mathrm{all}}$ is positive, as $\Delta b = b_2 - b_{\mathrm{all}} > 0$. However, as shown in Figure \ref{fig:dipo_z0}, for the cross-dipole among halos, and in Figure \ref{fig:dipo_z0_cdm}, for the cross-correlation between halos and CDM particles (the latter in redshift space as well), we conclude that no detection can be claimed with this pencil-beam light cone. Very similar results were obtained for the other two redshift samples $z_1 \approx 2.29$ and $z_2 \approx 2.69$.

Still, we point out the possibility of exploring optimal weighting schemes to enhance the signal in the light of the work carried out by \citet{castorina2019}. Lastly, the advantages of employing different tracers, coupled with the low densities of the catalogues, as well as the need for robust statistics, suggests the use of optimal weights for a more efficient combination of tracers \citep{abramo2015,montero2020}.

\section{Conclusions}\label{sec:conclusions}

We have performed a power spectrum multipole analysis on data from a light cone generated from a fully relativistic $N$-body simulation. We focused on the dipole signal in the cross-correlation between different dark matter halo subpopulations, which is a purely relativistic (non-Newtonian) effect. The simulation was generated by the \textit{gevolution} code, which employs a novel ray-tracing method to connect the halos with the observer, and which is capable of incorporating all relevant general relativistic effects on cosmologically-relevant distance scales. We showed in detail how the survey window function and quantities such as the evolution bias can be estimated on the past light cone, allowing a rigorous comparison with gauge-invariant theoretical calculations at linear order.

Similar studies of relativistic observables in simulations have been made in the past. For example, \citet{breton2019} and \citet{beutler2020} used the full-sky RayGal simulation, which is limited to the redshift range of $0.05 < z < 0.465$, with an effective $\bar{z} \sim 0.341$. While the simulation that we based our study on in principle covers the redshift range $0 \le z \le 7.1$, our analysis focused on a particular high redshift bin in the range $1.7 \le z \le 2.9$, which we further sliced into three different samples to keep the redshift evolution well controlled, covering a sky fraction of only $f_{\rm sky} = 0.01$. This is a similar sky area to the overlap region between different LSS tracers (luminous red galaxies and emission line galaxies) in the multi-tracer analysis of the final eBOSS data \citep[see Table 2 of][]{zhao2020}, although these data are from lower redshift, $z \sim 1$.

While we were able to robustly test our analysis methods using these simulated data, no conclusive detection of the dipole signature was possible due to the limited volume of the redshift bin, a challenge that is of paramount importance for current surveys too. \citet{beutler2020} studied the possibility of subtracting various contributions to the total signal in order to isolate the Doppler contribution and remove sample variance. Since the Doppler term is expected to increase in amplitude with redshift, one could also consider developing an optimal weighting scheme to enhance the signal and improve the prospects of detection \citep{castorina2019}. We leave this, and other schemes \citep{abramo2015, abramo2017, montero2020} to enhance detectability of the signal, to be explored in future work\CG{,} however. 

We did not incorporate wide-angle effects in our modelling, as they are not relevant for the solid angle and redshift range of our analysis. \CG{Nonetheless,} a careful account of these effects should also be explored in the context of wider survey areas, particularly in the case of future surveys such as Euclid, LSST, and SKA, which are expected to cover an appreciable fraction of the sky. 

Similarly, integrated effects (e.g. lensing), while fully included in our mock data, were neglected in our analytical model, but are known to impact large angular scales. Despite the Doppler term being the largest contribution to the relativistic effects for our particular setup, non-local terms should also be modelled and properly included for analyses that go to larger scales.

In this paper, we have limited our analysis to a single high-redshift bin with a relatively narrow survey area, and have pursued only a limited set of observables, i.e. the multipoles of the relativistic power spectrum. In future work, we will relax these limitations by moving to larger survey volumes more representative of the next generation of large-scale structure surveys, while also including wide-angle and integrated effects, and extending our analysis to two-point correlation functions and multipoles of the relativistic bispectrum.

\section*{Acknowledgements}

We thank David Alonso for making CUTE public. CG thanks Rodrigo Voivodic, Florian Beutler, and Michel-Andrés Breton for comments and fruitful discussions. In particular, she owes special thanks to Marcos Lima for all the support during the first stages of this work. She would also like to thank Queen Mary University of London for hospitality. CG was supported by QMUL (PHY2420B) and FAPESP (\href{https://bv.fapesp.br/pt/bolsas/180562/tecnicas-estatisticas-para-levantamentos-futuros-extraindo-fisica-primordial-da-estrutura-em-larga/?q=2018/10396-2}{2018/10396-2}) grants. JA acknowledges funding by STFC Consolidated Grant ST/P000592/1. CC is supported by STFC Consolidated Grant ST/P000592/1. LRA acknowledges financial support from CNPq (306696/2018-5) and FAPESP (\href{https://bv.fapesp.br/pt/auxilios/92620/explorando-novas-fronteiras-com-levantamentos-de-galaxias/?q=2015/17199-0}{2015/17199-0} and \href{https://bv.fapesp.br/pt/auxilios/101444/estruturas-em-grandes-escalas-no-universo-local-com-o-levantamento-s-plus/?q=2018/04683-9}{2018/04683-9}). This work was supported by a grant from the Swiss National Supercomputing Centre (CSCS) under project ID s710.

\section*{Data availability}
The data used in this work were acquired from the Swiss National Supercomputing Centre (project ID s710) and are available from the corresponding author upon a reasonable requests.

\paragraph*{Carbon footprint:} In this work, we reused existing data from a simulation that consumed about 8000 kWh of electrical energy. This has an estimated impact of 1600 kg $\mathrm{CO}_2$ when we use the conversion factor of 0.2 kg $\mathrm{CO}_2~\mathrm{kWh}^{-1}$ suggested by \citet{VuarnozJusselme2018} (see Table 2 therein, assuming Swiss mix). The additional energy used during the numerical analysis of the data is insignificant in comparison. This work also included a round trip São Paulo $\leftrightarrow$ London economy flight, emitting approximately 900 kg $\mathrm{CO}_2$\footnote{\href{https://www.icao.int/environmental-protection/CarbonOffset/Pages/default.aspx}{ICAO Carbon Emissions Calculator}, 25 August (2020).}.


\bibliographystyle{mnras}
\bibliography{refs.bib}


\appendix

\section{Halo properties}\label{ap:halos}
\subsection{Halo mass function}\label{ap:hmf}
For completeness\CG{,} we computed the mass function of our full halo sample between $z = 0.05$ and $z=0.465$ (effective redshift $\bar{z} = 0.34$) for comparison with the full-sky RayGal simulation\footnote{The RayGal simulation is contained within $0.05 < z < 0.465$\CG{. Their} effective redshift is the same as the low redshift considered in this section\CG{, for comparison}.} employed in the analysis of \citet{breton2019}. The halo mass function describes the probability of having a comoving number density of halos at redshift $z$ in the range $[\ln M,\ln M+d\ln M]$:
\begin{equation}
    \frac{\partial\bar{n}(M,z)}{\partial\ln M} = \frac{\bar{\rho}_{m,0}}{M} f(\sigma) \frac{\partial\ln \sigma^{-1}}{\partial\ln M}, \label{massfunc}
\end{equation}with $\bar{\rho}_{m,0}$ the comoving background matter density today, $f(\sigma)$ the multiplicity function, and $\sigma$ the overdensity variance smoothed in a sphere of radius $R$. The multiplicity function can be computed analytically from the spherical collapse model \citep{press-schechter} or the ellipsoidal collapse \citep{sheth1999}, or from numerical fits \citep{tinker2008}.

The RayGal simulation consists of a set of high-resolution Newtonian $N$-body simulations, whose halos have been ray traced to the redshift\CG{-}space position, rendering them with almost all properties of our halos. The RayGal light cone was built from 300 snapshots to avoid time discretisation effects.

\begin{figure}
    \centering
    \includegraphics[width=\columnwidth]{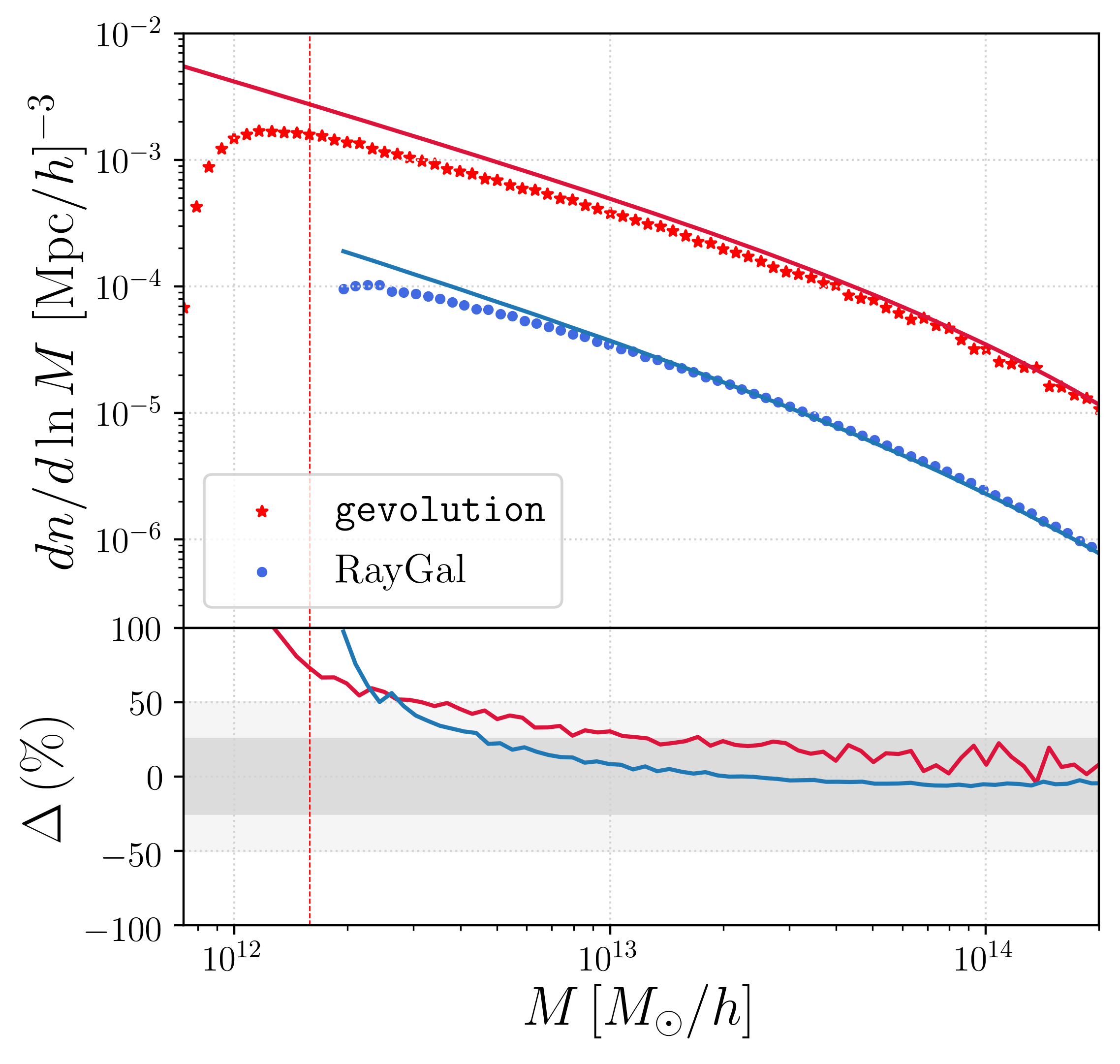}
    \caption{Comparison between the halo mass function of the catalogue employed in this analysis (\textit{stars}), with the one from the full-sky RayGal simulation (\textit{dots}) from \citet{breton2019}, with both catalogues in real space. Solid curves correspond to the Tinker mass function fit \citep{tinker2008}, while the bottom panel shows the relative difference between the fit prediction and the mass function computed from the simulations. Vertical dashed line corresponds to the limit of \textit{gevolution} halos with at least 600 particles. The cosmological parameters in the RayGal simulation that differ from ours are $h = 0.72$, $A_s = 2.431 \times 10^{-9}, \Omega_m = 0.257$, $T_{\mathrm{cmb}}= 2.726$. We stress that the RayGal mass function was multiplied by a 0.1 factor for a cleaner \CG{visualisation}, as the values were very similar. Differences between the Tinker and Sheth-Tormen \citep{sheth1999} mass functions were minor, so we only present the former.}
    \label{fig:raygalcomp}
\end{figure}

Our analysis was based on halo masses $M_{200b}\equiv M$ defined within the density thresholds of $\Delta = 200$, whose correspondence with the parameters fit of the Tinker mass function \citep[][solid lines in Figure \ref{fig:raygalcomp}]{tinker2008} is straightforward. In \citet{corasaniti2018}, RayGal halo mass functions were computed from the snapshots and were based on the Sheth-Tormen \citep{sheth1999} fit, with the halo identified with the spherical overdensity (SO) method, and thus the halos are more closely connected to the ellipsoidal collapse employed in the Sheth-Tormen fit \citep*{desjacques2018}.

However, for the RayGal light cone, halos were identified via a friends-of-friends (FOF) algorithm, just like in our catalogue. In \citet{smith2017}, differences with numerical fits seem at the low-mass end are also present, and they conclude that such discrepancies are associated with the comparison between different halo finder methods (SO and FOF). They computed the mass function using the SO correspondent, and just as in our case, found the same behaviour at low masses. In the catalogue employed in our analysis, $\sim 1.1 \times 10^6$ halos with $M_{\mathrm{vir}} \in [0.518,4.862] \times 10^{12} \,\,(M_{\odot}/h)$ were discarded for having their respective $M_{200b}$ null. As pointed out in \citet{smith2017}, small overdensities in large FOF groups might be identified as part of the larger group, leading to a lack of such structures.

Such discrepancies are important if one wishes to paint galaxies to the halos via, e.g., a halo occupation distribution. For our current purposes, the lack of a proper function to describe the light cone halo mass function impacted only our ability to predict the linear halo bias (see Section \ref{ap:bias}), and thus did not pose an issue for the analysis.

\subsection{Halo bias}\label{ap:bias}
\CG{As mentioned in the previous section, the} halo mass function describes the fraction of matter inside dark matter halos. So in order to obtain the correct halo statistics, we must account for their position in space. The halo bias, which is defined by the ratio of the halo power spectrum, $P_{hh}(k)$, to the linear dark matter power spectrum, $P_{\mathrm{lin}}(k)$ \citep{tinker2010},
\begin{equation}
    b^2(k) = \frac{P_{hh}(k)}{P_{\mathrm{lin}}(k)},\label{eq:halo_bias}
\end{equation}is best understood within the context of the peak-background split (PBS), where the long-wavelength modes \CG{enhance} the probability of forming halos by decreasing the threshold $\delta_c(z=0) = 1.686$ for overdensities \CG{located} at the peak of large-scale (background) fluctuations. It can be either derived from analytical mass functions, giving the Press-Schechter and Sheth-Tormen halo biases, or from equation (\ref{eq:halo_bias}) via numerical simulations. From \citet{tinker2010}, the bias is given by
\begin{equation}
    b(\nu) =1 - A\frac{\nu^a}{\nu^a + \delta_c^a} + B\nu^b + C\nu^c, \label{eq:tinkerbias}
\end{equation}where $\nu = \delta_c/\sigma$ and $A$, $B$, $C$, $a$, $b$, and $c$ are parameters fitted from simulations, depending on the matter perturbations at virialisation\CG{,} which is chosen to be $\Delta = 200$. This phenomenological fit proved to be unsatisfactory for our halo samples, for the reasons described in Section \ref{ap:hmf}. 

We proceeded then with the definition of equation (\ref{eq:halo_bias}) and employed a polynomial fit 
\begin{equation}
    b^2(k) = b_1^2 + b_2^2 k,\label{eq:pol_fit}
\end{equation}
considering the linear term as the fit for the linear halo biases, neglecting the scale dependence emerging from nonlinear effects in the power spectrum. Notice that the estimated spectra $P_{hh}$ employed in this fit are for the halos in real space. The results are shown in Table \ref{tab:bias_eqbin} for each halo sample, where we use different methods to obtain the matter power. In one case (PS) we used the theoretical power spectrum obtained from the CLASS Boltzmann solver \citep{blas2011} as \CG{the} denominator in equation (\ref{eq:halo_bias}). In a second case (CDM), we computed the biases using the real-space matter power spectrum computed from the cold dark matter particle ensemble instead of CLASS. As we can see, it differs by $\sim 9\%$ from the Tinker value, which we believe comes from fluctuations of the estimator itself. Despite that, we use this value to consistently compare the halo monopole obtained from the Legendre expansion with the real-space CDM spectrum in Figure \ref{fig:mono_z0}.

\begin{table}
    \small
    \centering
    \caption{Linear biases for the halo samples considered in this work, estimated with different methods. The Tinker bias is computed from the fit of \citet{tinker2010}, equation (\ref{eq:tinkerbias}). The biases in the next two columns were obtained via the polynomial fit of equation (\ref{eq:pol_fit}), with the linear power spectrum from CLASS (PS) or with the power spectrum estimated from the real-space cold dark matter particles (CDM), with the former estimate differing from the Tinker bias by $\sim 3-5\%$. The final column shows the bias parameters estimated from the correlation function, differing from the Tinker bias by $\sim 6\%$. Larger discrepancies for the CDM bias estimates with respect to the others will be due to fluctuations coming from the estimator.}
    \begin{tabular}{l c c c c c}
        \hline
          & Mass            &  Bias     & Bias  & Bias & Bias \\
          & $[M_{\odot}/h]$ & (Tinker)  & (PS) & (CDM) & (CF) \\
        \hline
        \multicolumn{6}{c}{$\bar{z} = 1.89$} \\
        \hline
        All & $4.4 \times 10^{12}$ & 2.881 & 2.927 & 3.076 & 2.809\\
        H$_{0}$  & $1.9\times 10^{12}$ & 2.273 & 2.551 & 2.683 & 2.437\\
        H$_{1}$  & $2.8\times 10^{12}$ & 2.540 & 2.758 & 2.897 & 2.652\\
        H$_{2}$  & $8.5\times 10^{12}$ & 3.539 & 3.477 & 3.651 & 3.349\\
        
        \hline
        \multicolumn{6}{c}{$\bar{z} = 2.29$} \\
        \hline
        All & $3.9 \times 10^{12}$ & 3.473 & 3.469 & 3.749 & 3.196\\
        H$_{0}$  & $1.8\times 10^{12}$ & 2.803 & 3.020 & 3.280 & 2.772\\
        H$_{1}$  & $2.7\times 10^{12}$ & 3.114 & 3.270 & 3.546 & 2.993\\
        H$_{2}$  & $7.1\times 10^{13}$ & 4.212 & 4.154 & 4.449 & 3.827\\
        \hline
           
        \multicolumn{6}{c}{$\bar{z} = 2.69$} \\
        \hline
        All & $3.4 \times 10^{12}$ & 4.140 & 4.214 & 4.455 & 3.959\\
        H$_{0}$  & $1.8\times 10^{12}$ & 3.414 & 3.735 & 3.950 & 3.613\\
        H$_{1}$  & $2.5\times 10^{12}$ & 3.765 & 4.006 & 4.237 & 3.891\\
        H$_{2}$  & $5.9\times 10^{13}$ & 4.955 & 4.932 & 5.211 & 4.468\\
        \hline
    \end{tabular}
    \label{tab:bias_eqbin}
\end{table}

In order to verify the consistency of the method and the possible impacts of the window function (for the CLASS case) and estimator (for the CDM) in the bias estimation, we also tested the approach of \citet{breton2019}: here the linear bias was computed by fitting a constant function to the ratio
\begin{equation}
    b = \sqrt{\frac{\xi_{hh}^{\ell=0}}{\xi_{0}}},\label{eq:mono_ratio}
\end{equation}where $\xi_0$ is the monopole of the matter autocorrelation function, computed from
\begin{equation}
    \xi_0(x) = \frac{1}{2\pi^2} \int \mathrm{d}k \, k^2 j_0(kx)P^{(r)}(k),
\end{equation}and $\xi_{hh}^{\ell=0}$ is monopole of the halo-halo \CG{autocorrelation}, computed from the real\CG{-}space catalogues with \texttt{CUTE}\footnote{\href{https://github.com/damonge/CUTE/}{https://github.com/damonge/CUTE/}} \citep{alonso2012}. The real\CG{-}space linear matter power spectrum\footnote{Differences of using the nonlinear matter power spectrum were below the per cent level.} was obtained from CLASS with the input parameters of the simulation. 

This method is not perfect though: we observed a shift in the BAO peak scale for the halo samples if compared to the theoretical prediction. Also, we limited ourselves to the range $ 28 < r_{\mathrm{fit}} < 68$ in units of Mpc/$h$. Although time consuming, this method is naturally safe from the mode coupling induced by the window function. The results from this fit are shown in Table \ref{tab:bias_eqbin}. Using the fits from equation (\ref{eq:pol_fit}) or (\ref{eq:mono_ratio}) did not lead to significant differences to the overall results, with both methods being equally tantamount. 

\subsection{Evolution bias}\label{ap:evolbias}

\begin{figure*}
    \centering
    \includegraphics[width=\textwidth]{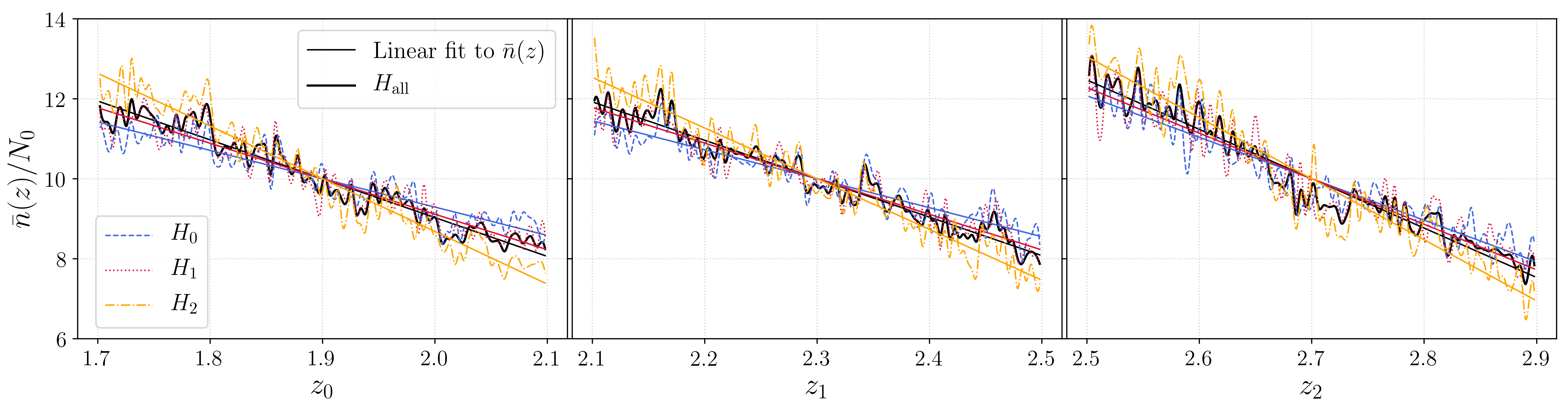}
    \caption{The comoving number density of halos, normalised by the total amount $N_0$ of objects inside each mass bin. Solid lines represent the linear fit of equation (\ref{eq:linear_nbar}), whereas dashed lines represent the true $\bar{n}(z)$ of the simulation, computed by dividing each redshift slice into 100 bins to capture the redshift evolution. The smooth curves are the result of a cubic spline interpolation, for visual reasons.}
    \label{fig:nbarNorm_eq}
\end{figure*}

\begin{figure*}
    \centering
    \includegraphics[width=\textwidth]{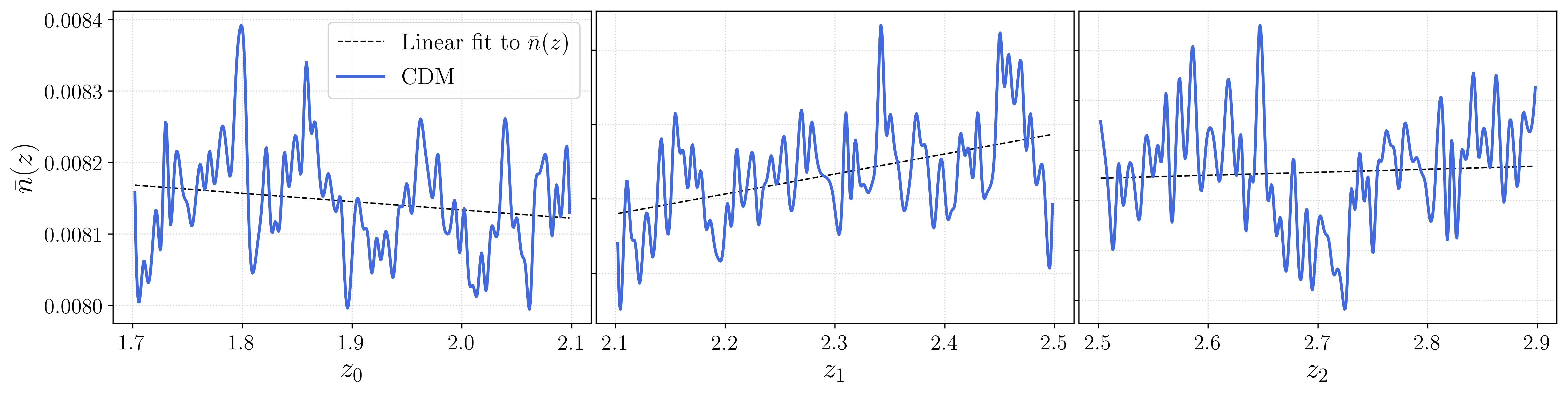}
    \caption{The comoving number density $\bar{n}(z)$ of cold dark matter (CDM) particles. Dashed lines represent the linear fit of equation (\ref{eq:linear_nbar}). As in Figure \ref{fig:nbarNorm_eq}, the true $\bar{n}(z)$ extracted from the simulation (solid curve) was computed using 100 bins inside each redshift slice considered for analysis, with the smoothness being a result of a cubic spline interpolation.}
    \label{fig:nbar_cdm}
\end{figure*}

The evolution bias of LSS tracers quantifies the intrinsic variation in the number of sources in the Universe, and thus gives information about the time evolution of tracers. It is defined in equation (\ref{eq:evolbias}), and depends on the comoving number density of sources in real space $\bar{n}$\footnote{The evolution bias can also be defined in terms of the physical number density. In this case, one must account for the fact that, instead of $b_e = 0$ for cold dark matter particles, $b_e = 3$.}. \CG{In the case of dark matter halos, coming from a simulation, this parameter is completely faithful to the intrinsic cosmological variations of the mean number density of halos, as the underlying dark matter distribution is known and the comoving number density of halos is complete (i.e., all halos that were supposed to be found are included in the catalogue).}

In Figure \ref{fig:nbarNorm_eq}, we show the comoving number density of real\CG{-}space halos, inside each redshift bin considered in the main analysis, normalised by the total number of halos for each mass bin defined for $\bar{n}_0 \approx \bar{n}_1 \approx \bar{n}_2$. As one can see, there is a large variation explained by the fact that more massive halos are more common at lower redshifts, which can be explicitly seem by the slope of the curves. This intrinsic variation is captured by the evolution bias parameter $b_e$, and is a major parameter entering the relativistic corrections. 

Analogously, we also show in Figure \ref{fig:nbar_cdm} the comoving number density of the cold dark matter (CDM) particles, in real space, used to build the halo catalogue. This was used to derive, in the same way as done for the halos, the evolution biases for the particles within each redshift slice. We obtained $b_e^{\mathrm{CDM}} = \lbrace 0.041, -0.109, -0.014 \rbrace$ as best fit, respectively for the three redshift bins $\bar{z} = \lbrace 1.89, 2.29, 2.69 \rbrace$. This is consistent with what we expect for the CDM particles because the comoving number density is constant, the true evolution bias vanishes. However, large-scale density gradients due to matter perturbations will lead to a non-zero best fit within any finite volume.

Following \citet{beutler2020}, we fit a linear function to the (unnormalised) comoving number density,
\begin{equation}
    \bar{n}(z) = a+bz,\label{eq:linear_nbar}
\end{equation}which leads to the analytical expression for $b_e$,
\begin{equation}
    b_e(z) = \frac{c + 1}{c - z} - 1, \label{eq:evolbias_fit}
\end{equation}where $c \equiv a/(-b)$.

Notice that, even though the comoving mean number density is the same for all the three samples $H_0$, $H_1$ and $H_2$, the different evolution with redshift between the different halo populations, defined by different halo masses, leads to distinct evolution biases, as can be seen in Figure \ref{fig:evolbias_eq}.

\begin{table}
    \small
    \centering
    \caption{Parameters entering equations (\ref{eq:linear_nbar}) and (\ref{eq:evolbias_fit}) to derive the evolution bias of the halo samples.}
    \begin{tabular}{l c c c c }
        \hline
          & $a$ & $b$ & $c$ & $b_e(\bar{z})$\\
        \hline
        \multicolumn{5}{c}{$\bar{z} = 1.89$} \\
        \hline
        All & 20.105 & -6.863 & 2.929 & 2.781\\
        H$_{0}$  & 5.521 & -1.668 & 3.310 & 2.035\\
        H$_{1}$  & 6.337 & -2.093 & 3.027 & 2.542\\
        H$_{2}$  & 8.247 & -3.102 & 2.658 & 3.761\\
        \hline
        \multicolumn{5}{c}{$\bar{z} = 2.29$} \\
        \hline
        All & 14.990 & -4.488 & 3.340 & 3.132 \\
        H$_{0}$  & 4.153 & -1.128 & 3.680 & 2.367 \\
        H$_{1}$  & 4.740 & -1.385 & 3.422 & 2.905 \\
        H$_{2}$  & 6.098 & -1.974 & 3.088 & 4.120 \\
        \hline
        \multicolumn{5}{c}{$\bar{z} = 2.69$} \\
        \hline
        All & 12.763 & -3.636 & 3.510 & 4.500 \\
        H$_{0}$ & 3.728  & -1.017 & 3.664 & 3.788 \\
        H$_{1}$ & 4.000 & -1.118 & 3.578 & 4.153 \\
        H$_{2}$ & 5.035 & -1.501 & 3.355 & 5.551 \\
        \hline
    \end{tabular}
    \label{tab:nbarz_fit_eq}
\end{table}

\section{Window function}\label{ap:window}

In this appendix we describe how the window function is obtained. It is employed to compute the observed power spectra multipoles from the theoretical predictions\CG{,} for comparative purposes\CG{,} and imparts substantial effects on large scales and on the odd multipole moments. Therefore, its inclusion is mandatory.

We begin by recalling that the observed density field $\hat{\delta}$ is given by 
\begin{equation}
    \hat{\delta}(\boldsymbol{x}) = W(\boldsymbol{x}) \delta(\boldsymbol{x}),
\end{equation}where $\delta(\boldsymbol{x})$ is the true underlying density field and $W(\boldsymbol{x}) = w(\boldsymbol{x}) \bar{n}(\boldsymbol{x})$ accounts for the survey geometry and local weighting $w$ scheme \citep{FKP}. Therefore, the observed correlation function is given by
\begin{equation}
    \hat{\xi}(\boldsymbol{s}_1, \boldsymbol{s}_2) = W(\boldsymbol{s}_1)W(\boldsymbol{s}_2) \xi(\boldsymbol{s}_1,\boldsymbol{s}_2).\label{eq:correlation}
\end{equation}Notice that we can write $\boldsymbol{s}_2 = \boldsymbol{s}_1 + \boldsymbol{s}$, where $\boldsymbol{s}$ is the pair separation, so that the correlation function may also be written as $\xi(\boldsymbol{s}_1, \boldsymbol{s})$ (see e.g. Figure \ref{fig:config}).

In Fourier space we obtain the well-known convolution result for the overdensity field, 
\begin{equation}
    \hat{\delta}(\boldsymbol{k}) = \int \frac{\mathrm{d}^3k'}{(2\pi)^3} W(\boldsymbol{k}-\boldsymbol{k}') \delta(\boldsymbol{k}'), 
\end{equation}yielding the three-dimensional observed power spectrum:
\begin{equation}
\begin{aligned}
    \hat{P}(\boldsymbol{k}) &= \int \frac{\mathrm{d}^3q}{(2\pi)^3}  |W(\boldsymbol{k}-\boldsymbol{q})|^2 P(\boldsymbol{q}),\\
    &= \int \frac{\mathrm{d}^3q}{(2\pi)^3}  |W(\boldsymbol{q})|^2 P(\boldsymbol{k}-\boldsymbol{q})\label{eq:conv_power},
\end{aligned}
\end{equation}where we made use of the fact that $\delta$ and $W$ are real quantities.

One possible way to compare theory and estimates is to deconvolve the survey window from $\hat{P}$; however, since convolution in Fourier space destroys information, the deconvolution of the window is an attempt to recover this intrinsic information loss in the signal analysis. The standard procedure \citep{beutler2017,wilson2017} consists, instead, in computing the multipoles of $|W(\boldsymbol{q})|^2$ to convolve the theoretical power spectrum to obtain $\hat{P}$, where 
\begin{equation}
    |W(\boldsymbol{q})|^2 =  \int \mathrm{d}^3s\, \mathrm{e}^{-i\boldsymbol{q} \cdot \boldsymbol{s}} W^2(\boldsymbol{s}),
\end{equation}and
\begin{equation}
    W^2(\boldsymbol{s}) \equiv \int \mathrm{d}^3s_1 \,W(\boldsymbol{s}_1)  W(\boldsymbol{s}_1+\boldsymbol{s}).
\end{equation}

We shall write $W^2(\boldsymbol{s}) \equiv Q(\boldsymbol{s})$ and $|W(\boldsymbol{k})|^2 = Q(\boldsymbol{k})$. Notice that this depends on the local LOS, which is taken to be $\boldsymbol{s}_1$ (end-point LOS) in the case of the YBS multipoles estimator. Hence, the multipoles of the ``window function'', with respect to a LOS $\hat{\boldsymbol{d}} = \hat{\boldsymbol{s}}_1$, are given by
\begin{equation}
    Q_{\ell}(s) = \frac{2\ell+1}{4\pi} \int \mathrm{d}^3 d\int \mathrm{d}\Omega_s \,\, Q(\boldsymbol{s},\boldsymbol{d}) \mathcal{P}_{\ell}(\hat{\boldsymbol{s}}\cdot\hat{\boldsymbol{d}})\label{eq:window_config}
\end{equation}in configuration space\CG{,} and
\begin{equation}
    Q_{\ell}(k) = \frac{2\ell+1}{4\pi} \int \mathrm{d}^3 d\int \mathrm{d}\Omega_k \,\, Q(\boldsymbol{k},\boldsymbol{d}) \mathcal{P}_{\ell}(\hat{\boldsymbol{k}}\cdot\hat{\boldsymbol{d}})
\end{equation}in Fourier space. The integrals on $\mathrm{d}\Omega_s$ and $\CG{\mathrm{d}}\Omega_k$ run over the angles between, respectively, $\hat{\boldsymbol{s}}$ and $\hat{\boldsymbol{k}}$ with the LOS $\hat{\boldsymbol{d}}$: $\int \CG{\mathrm{d}}\Omega = \int_0^{\pi} \CG{\mathrm{d}}\theta \sin\theta \int_0^{2\pi} \CG{\mathrm{d}}\varphi$. After the integration over all angles, $Q_{\ell}(s)$ can be obtained by the final integration over all possible LOS. Because our ``survey'' geometry is well-behaved\footnote{For surveys whose angular selection is too complicated, precluding an analytical derivation of the window function, one can compute it from FFTs of a random catalogue, as we discuss further, or from a random-random pair count \citep{beutler2017,wilson2017}; this latter option, however, can be too time consuming for surveys that require a large number of random objects\CG{. To circumvent this issue,} \citet{breton2020} proposed a similar approach to \CG{semi-analytically} compute the pair counts with high precision and without much computational effort.}, consisting of a simple angular selection delimited by the light cone opening angle (which sets the upper limit in the $\theta$ integral\CG{)}, $Q_{\ell}(s)$ can be obtained semi-analytically by considering the halos radial selection function (shown in Figure \ref{fig:nbarNorm_eq}), and performing a plain angular integration, with the radial integral limited to the redshift range of the sample through the proper inclusion of Heaviside step functions.

\begin{figure}
    \centering
    \includegraphics[width=\columnwidth]{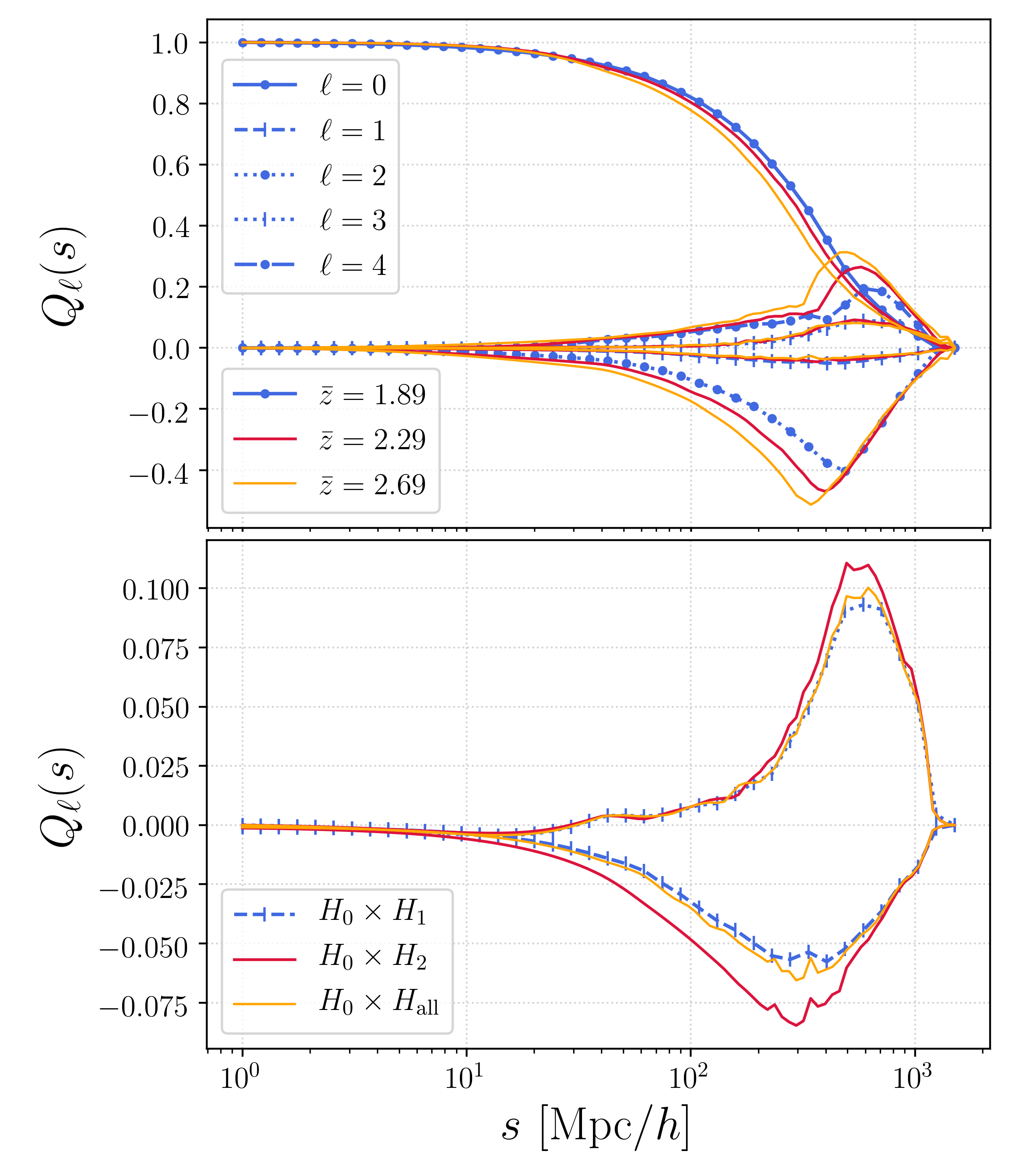}
    \caption{Multipoles of the window function computed semi-analytically from the \CG{radial and angular distribution of the samples}. The top panel shows the window for the catalogues of the halo population $H_0$ at the redshift slices considered for analysis (different colours), with different line styles representing different multipoles: dots depict the even, while vertical bars the odd ones. Differences between the tracers as a result of different selection functions are negligible, and thus we only show the autocorrelation of the $H_0$; however, the cross-correlation among tracers has a larger impact for the odd multipole moments, as shown in the bottom panel for a fixed redshift ($\bar{z} = 1.89$).}
    \label{fig:theory_window}
\end{figure}

With the theoretical window function, obtained from its definition in equation (\ref{eq:window_config}) and shown in Figure \ref{fig:theory_window}, differences between halo populations as a result of different selection functions were below the 2\% level for the autocorrelation case, with the same behaviour being observed \CG{for} the windows \CG{estimated} from the random catalogues. In contrast, differences between redshift bins are more relevant, as can be seen in the \CG{upper} panel (\CG{different colours}) of Figure \ref{fig:theory_window}, and must be fully included in any analysis, \CG{whereas different selection functions impact the odd multipoles of the window function (Figure \ref{fig:theory_window}, bottom panel)}.

If we plug in the expressions for $Q(\boldsymbol{k},\boldsymbol{d})$ and $Q(\boldsymbol{s},\boldsymbol{d})$, just as in the local power spectrum case, we see that to estimate the multipoles $Q_{\ell}(k)$ and $Q_{\ell}(s)$ we just apply the usual power spectrum and correlation function estimators. Since the explicit convolution of equation (\ref{eq:conv_power}) is computationally expensive, and so is the computation of $Q_{\ell}(s)$ directly from the random pair correlation, as the survey window function (random catalogue) contains $10^8$ particles to completely fill the survey region, one possibility is to compute the power spectrum multipoles of the random catalogues. \CG{With this strategy, $Q_{\ell}(k)$ is quickly obtained by means of FFTs}, which are then taken to configuration space \CG{\citep{beutler2019} for the proper convolution}. 

We compared this approach with the theoretical window and found disagreements for large scales due to possible instabilities in the Hankel transform, and the limited $k$ range and fluctuations from the FFT estimator that made the window function very noisy for large $s$. For this reason, we opted to analyse the impact of the window by using the semi-analytical result.

\begin{figure}
    \centering
    \includegraphics[width=\columnwidth]{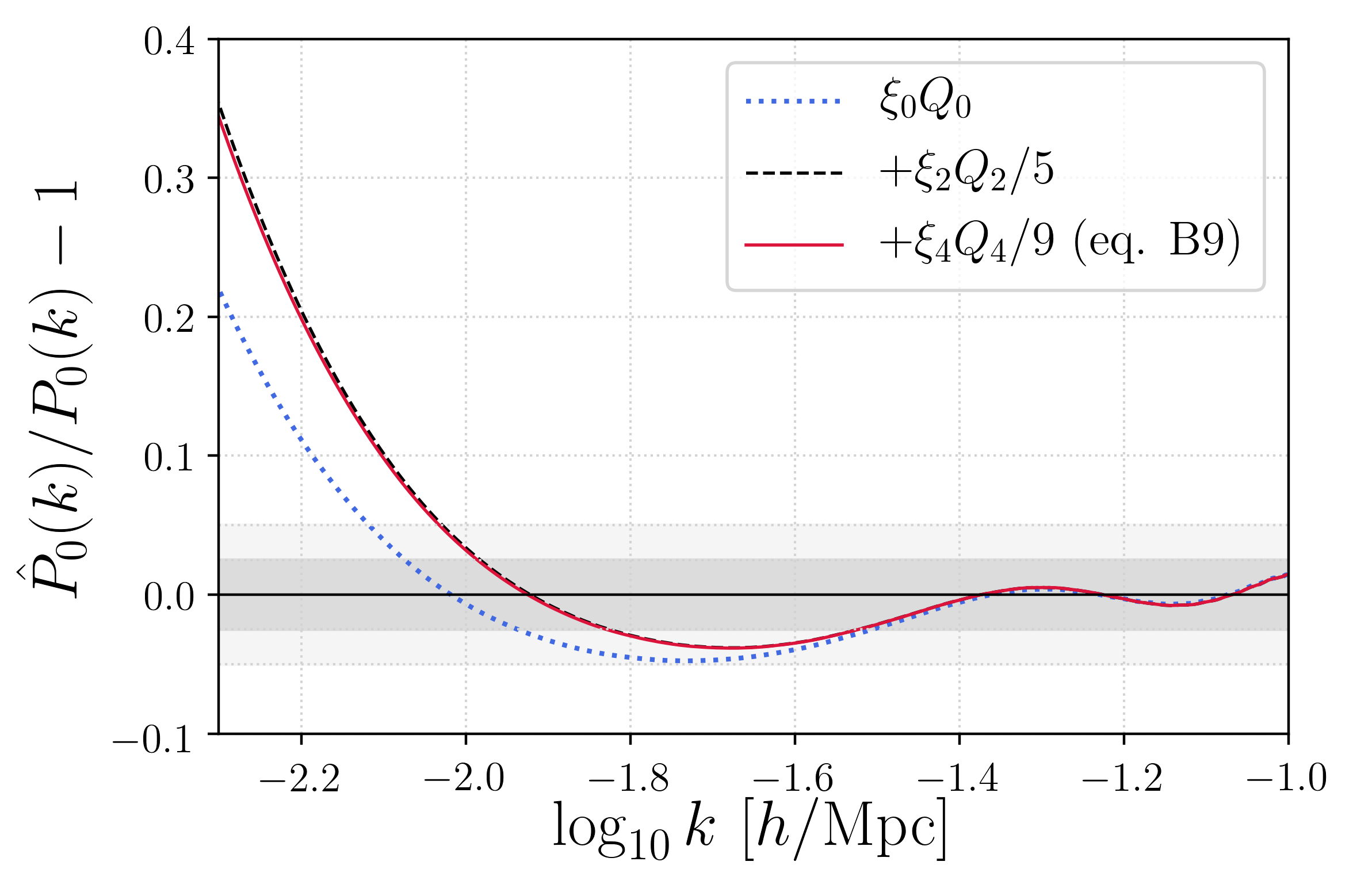}
    \caption{Relative difference between the \CG{convolved $\hat{P}_0$} and the linear power spectrum monopole obtained from the Newtonian redshift-space \CG{prediction} (as described in equations \ref{eq:legendre}, \ref{eq:power_multipole_coeff} and \ref{eq:kaiser_mono}). The convolved theory was obtained via equation (\ref{eq:conv_mono}), for the cases where there is no leakage of the quadrupole and hexadecapole to the monopole (blue\CG{-}dashed curve), where we only account for the quadrupole contribution to the monopole (black\CG{-}dashed curve), that is, considering the first two terms in the right-hand side of equation (\ref{eq:hatxi0}), and given the full expression (red\CG{-}solid curve). Shaded regions corresponds to differences of 0.05 (light grey) and 0.025 (darker grey). The inclusion of the quadrupole becomes relevant for scales $k \lesssim 10^{-1.6} = 0.025$ $h/$Mpc, while the hexadecapole contribution is negligible.}
    \label{fig:convolution}
\end{figure}

From the straightforward product of equation (\ref{eq:correlation}), the Legendre expansion of (\ref{eq:correlation}) results in \citep{beutler2017,beutler2019,wilson2017,beutler2020}:
\begin{equation}
    \hat{\xi}_{0}(s) = \xi_0(s) Q_0(s) + \frac{1}{5}\xi_2(s) Q_2 + \frac{1}{9}\xi_4(s) Q_4(s) + \ldots , \label{eq:hatxi0}
\end{equation}and
\begin{align}
    \hat{\xi}_1(s) &= \xi_0(s) Q_1(s) + \xi_2(s) \left[\frac{2}{5} Q_1(s) + \frac{9}{35} Q_3(s)\right] \nonumber \\
    &{}\hspace{18mm} + \frac{4}{21} \xi_4(s)Q_3(s) + \ldots,\label{eq:hatxi1}
\end{align}and from an inverse Hankel (1D Fourier) transform of these equations we finally obtain the convolved power spectrum multipoles of equation (\ref{eq:conv_power}). For the monopole we have
\begin{equation}
    \hat{P}_0(k) = 4\pi \int s^2\, \mathrm{d}s \, j_{\ell}(ks) \, \hat{\xi}_0(s).\label{eq:conv_mono}
\end{equation}In Figure \ref{fig:convolution} we show the relative difference between what is obtained from equation (\ref{eq:conv_mono}) and the linear power spectrum monopole. The latter is computed from the Legendre expansion, with the Newtonian monopole coefficient $c_0$ shown in equation (\ref{eq:kaiser_mono}) and the linear real-space power spectrum extracted from CLASS. The former is computed in three different ways. First, we only include the first term in equation (\ref{eq:hatxi0}), which corresponds to the case where there is no coupling between higher order multipoles with $\ell = 0$. Then we account for the leakage of the quadrupole (second term in equation \ref{eq:hatxi0}), and of the hexadecapole (full equation \ref{eq:hatxi0}). As we can see, the inclusion of the quadrupole becomes relevant for scales $k \lesssim 0.025$ $h/$Mpc, while the hexadecapole contribution is negligible. For our range of scales, the differences between the convolved and the redshift-space theory (unconvolved) is of the order of 5\%. Thus, we do not see a large impact in face of our error bars (see Figure \ref{fig:mono_z0}). We stress that this behaviour is consistent with the BOSS DR12 anisotropic analysis \citep{beutler2017}.

Finally, following \citet{beutler2019} and \citet{beutler2020}, the convolved dipole is given by:
\begin{equation}
    \hat{P}_1(k) = -3i\int s^2\ \mathrm{d}s\ j_1(ks)\hat{\xi}_1(s) - i Q_1(k)\int s^2\ \mathrm{d}s\ \ \hat{\xi}_0(s).
\end{equation}This represents the leakage of even multipoles to the dipole, and must be \CG{accounted for} if one wishes to analyse the pure signal of the cross-dipole $P_1^{\alpha\beta}(k)$.

\bsp	
\label{lastpage}

\end{document}